\documentclass[3p,12pt]{article}
\usepackage[left=3cm,top=3cm,bottom=2.7cm,right=2.4cm]{geometry}
\usepackage{amsmath}
\usepackage{amssymb}
\usepackage{color}
\usepackage{graphicx}
\usepackage{epstopdf}
\usepackage{subcaption}
\usepackage{psfrag}
\usepackage{float}
\usepackage{hyperref}

\renewcommand{\div}{\nabla\cdot}
\def\bfq{{\bf q}}

\begin{document}
\title{The non-monotonicity of growth rate of viscous fingers\\ in heterogeneous porous media}
\maketitle
\begin{center}
\author{I.A.\,Starkov$^{1,2*}$, \and D.A.\,Pavlov${^2}$, \and S.B.\,Tikhomirov$^{2,3}$, \and F.L.\,Bakharev$^{2}$}\\[2ex]
\date{
$^{1}$All-Russian Research Institute of Fats,\\ Chernyakhovskij St.\,10., 191119, St.\,Petersburg, Russia\\
$^{2}$Saint-Petersburg State University,\\ Universitetskaya Emb.\,7/9, 199034, St.\,Petersburg, Russia\\
$^{3}$University of Duisburg-Essen,\\ Thea-Leymann-Str. 9, 45127, Essen, Germany\\
$^*$corresponding author; phone: +7(812)712-26-14,  e-mail: ferroelectrics@ya.ru\\[2ex]
}
\end{center}

\begin{abstract} 
The paper presents a stochastic analysis of the growth rate of viscous fingers in miscible displacement in a heterogeneous porous medium. The statistical parameters characterizing the permeability distribution of a reservoir vary over a wide range. The formation of fingers is provided by the mixing of different-viscosity fluids  --- water and polymer solution. The distribution functions of the growth rate of viscous fingers are numerically determined and visualized. Careful data processing reveals the non-monotonic nature of the dependence of the front end of the mixing zone on the correlation length of the permeability (describing the medium graininess) of the reservoir formation. It is demonstrated that an increase in graininess up to a certain value causes an expansion of the distribution shape and a shift of the distribution maximum to the region of higher velocities. In addition, an increase in the standard deviation of permeability leads to a slight change in the shape and characteristics of the density distribution of the growth rates of viscous fingers. The theoretical predictions within the framework of the transverse flow equilibrium approximation and the Koval model are contrasted with the numerically computed velocity distributions.
\end{abstract}


\section{Introduction}\label{intro}
The study of the mixing of two fluids has been the subject of extensive research in recent decades. In the area of contact between two mixing fluids, a so-called mixing zone is formed, the growth rate of which is determined by various factors. In the absence of external influence, mixing occurs by diffusion of one fluid into other. However, the presence of external forces can lead to a more active mixing process, which manifests itself in the form of thin viscous fingers. Famous examples of viscous fingers formation are the Hele-Shaw cell \cite{HS} and the Saffman-Taylor instabilities \cite{SafTay1, SafTay2}. The mixing of fluids with different viscosities is important in the oil industry because it can be triggered by a displacement process caused by pumping one fluid into a reservoir with another fluid. The issue is particularly acute in chemical oil recovery methods that involve the injection of highly viscous polymers as part of the process. Diffusion processes play a much smaller role in active mixing, and the size of the mixing zone has been found to depend linearly on time at high P\'eclet numbers and constant external forces (see, e.g.~\cite{nijjer, Linear}).

Unfortunately, despite a long history of research, the mechanics of viscous fingers formation in the mixing zone is poorly understood. As a result, the analysis of the mixing zone is often reduced to the study of average properties. In oil industry applications, it is important to be able to estimate the growth rate of the mixing zone and understand what key factors affect this rate (e.g., the influence of fluid viscosities, formation rock properties, injection rate). There are not many known theoretical results in this direction. In the case of gravitational mixing, we highlight the important works~\cite {otto2005,otto2006}. For the displacement of a more viscous fluid by a less viscous one, the development of~\cite{otto2005,otto2006} is carried out in \cite{yortsos2006}.

Due to the lack of an established theoretical framework, it is natural to consider numerical approaches to mixing zone growth under realistic parameters. Yet the challenge of implementing such numerical simulations remains an ongoing and independent problem. This is due, for example, to the existence of numerical diffusion and the requirement that the size of the grid elements be small compared to the finger width~\cite{chen1998part1, chen1998part2, ChuokeMeursPoel5, luo2017interactions, luo2018scaling}. Nevertheless, the problem of miscible displacement has been successfully solved with the help of numerical calculations for various application scenarios (see for example~\cite{mishra2008, chen2001, dewit2005, pramanik2016, sharma2021}).

In our opinion, the permeability heterogeneity should be considered as the most crucial parameter affecting the formation and growth rate of viscous fingers. Intuitively, the inhomogeneity in the permeability distribution should enhance the effect of the formation and propagation of viscous fingers. At the same time, when the correlation length is very small, it can be assumed that a permeability field can be effectively homogenized and the behavior of viscous fingers  resembles that of a uniform field. If the correlation length is large compared to the interwell spacing, the media should have little inhomogeneity in the region where viscous fingers develop, and the behavior should be also similar to that of a homogeneous structure. For intermediate values of correlation length, one should expect a nontrivial dependence on viscous fluid properties. This simple observation was one of the motivations for a detailed study of the dependence of viscous fingers velocity on the stochastic properties of the permeability fields. In particular, we expect that the dependence of the fingers growth rate is non-monotonic and that the velocity reaches a maximum for a certain correlation length. Understanding this maximum velocity is extremely important for generating pessimistic estimates for effective chemical flooding design in enhanced oil recovery methods.

In the present work we numerically investigate the miscible displacement of viscous fluids in heterogeneous porous media described by Darcy law and incompressible flow. A Gaussian field with certain properties represents the inhomogeneous permeability, as it is often done in geostatistics~\cite{CD99}. The most important characteristics of the permeability field for real-world applications in this case are dispersion and correlation length. Such types of models are often used in applied modeling of reservoirs. At the same time, analyzing the properties of solutions of partial differential equations in media characterized by a stochastic field is an interesting mathematical topic by itself, e.g.~in stochastic homogenization \cite{Koz79, PV79, GNO15} or traveling waves in stochastic media~\cite{NR09, Nol11}. 

\section{Statement of the problem}
The construction of a stochastic model for the viscous fingers growing through the polymer bank can be implemented with a single-phase miscible displacement in a linear reservoir corresponding to water-polymer flooding between two horizontal wells. 

\subsection{System of basic equations}
\begin{figure}[t]
    \centering
     \includegraphics[width=0.4\textwidth]{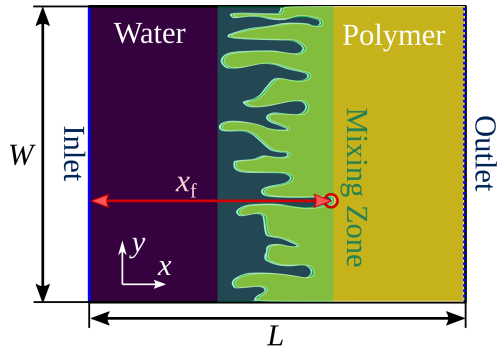}
        \caption{The schematic representation of a water-polymer flooding reservoir. The distance $x_\mathrm{f}$ corresponds to the tip of the longest water finger. The computational domain has dimensions $L = 40$\,m and $W = 31.4$\,m.
                }
\label{fig:slug}  
\end{figure}

The time-dependent incompressible flow generated by a miscible displacement process can be described using the Darcy law for porous media, mass balance, and continuity of pressure and flux conditions (also known as the Peaceman model) \cite{Peaceman1962}

\begin{equation}
\begin{aligned}
&\text{conservation of species}&&
\phi\cfrac{\partial  c}{\partial t}+ \div (\bfq\,c)=0,
\\
&\text{incompressibility condition}&& 
\div\bfq=0,
\\
  &\text{Darcy law}&&
  \bfq=-k\, m(c)\nabla p.
\end{aligned}
\label{eq:A}
\end{equation}
Here, $c$, $p$, $\bf q$ are the polymer concentration, pressure, and linear velocity, respectively. The mobility of the water phase $m(c)$ is inversely proportional to the viscosity $\mu(c)$, which is a known and monotonically increasing function of the concentration, $c$. We assume that gravity has no effect as our attention is focused on the two-dimensional flow in the horizontal plane. We model the random nature of oil reservoir structure, where the permeability is variable in space: $k(x,y)$. We consider the evolution of viscous fingers on the variance and correlation length of $k(x,y)$ assuming that the permeability is a stationary Gaussian field. The porosity of the porous medium, $\phi$, is considered constant. The system \eqref{eq:A} can be scrutinized in a linear geometry $(x,y)\in [0,L]\times [0,W]$ (see, Fig. \,\ref{fig:slug}) with the following boundary conditions

\begin{equation}
\begin{aligned}
    &\text{inlet}, x=0 && c=0, &&\bfq = q_0 \bf{e}_1, \vspace{0.2cm}\\
    &\text{outlet}, x=L &&  \cfrac{\partial c}{\partial x}=0, && p=0,\vspace{0.2cm}\\
    &\text{no-flow condition}, y=\{0,W\} &&  \cfrac{\partial c}{\partial x}=0, && \bfq\, {\bf e_2}=0,
\end{aligned}
\end{equation}
with ${\mathbf e_{1,2}}$ as unit vectors. The outflow condition is maintained at the outlet.
The reservoir is filled with the polymer solution of a certain concentration $c_\mathrm{max}$ before the water injection starts.

\section{Numerical framework\label{numerics}}
In order to obtain the statistics needed to build a stochastic model for each coarseness level describing heterogeneous porous media, a set of 200 maps with different spatial permeability profiles $k(x,y)$ but with the same scatter of values was generated using GPFlowSampling\,\cite{bib:gpflow}. The resulting maps were fed into a custom program utilizing the finite volume method for the water-polymer system of equations\,\eqref{eq:A}. The model is implemented on top of DuMu$^\mathrm{x}$\,\cite{bib:dumux11,bib:dumux20}, an open source framework for simulating flow and transport in porous medium. To generalize the simulation results, it is convenient to use relative units. Moreover, the growth rate of viscous fingers does not depend on the size of the computational domain, as shown in\,\cite{Linear}. In the following, all spatial dimensions are normalized to $L$ and time is measured in injected pore volumes (PV). 

\subsection{GPFlowSampling}
\begin{figure}[t!]
    \centering
    \begin{subfigure}{0.32\textwidth}
        \includegraphics[height=0.17\textheight]{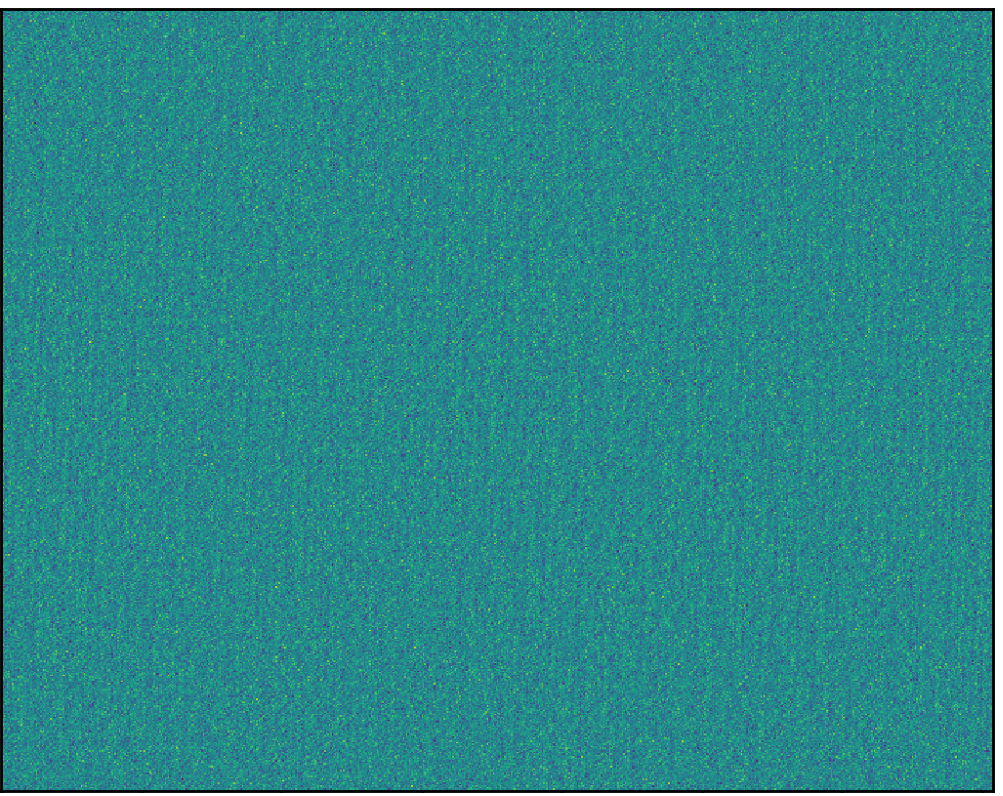}
        \caption{$l_\mathrm{c}=10^{-4}$.}
    \end{subfigure}
    \begin{subfigure}{0.32\textwidth}
        \includegraphics[height=0.17\textheight]{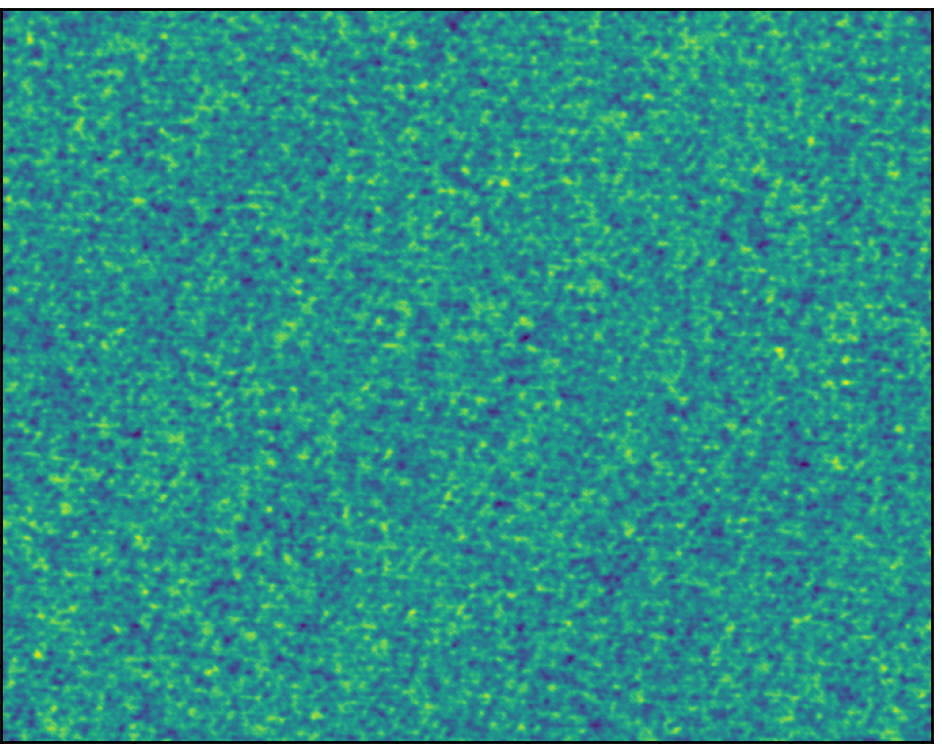}
        \caption{$l_\mathrm{c}=5\cdot10^{-3}$.}
    \end{subfigure}
    \begin{subfigure}{0.32\textwidth}
        \includegraphics[height=0.17\textheight]{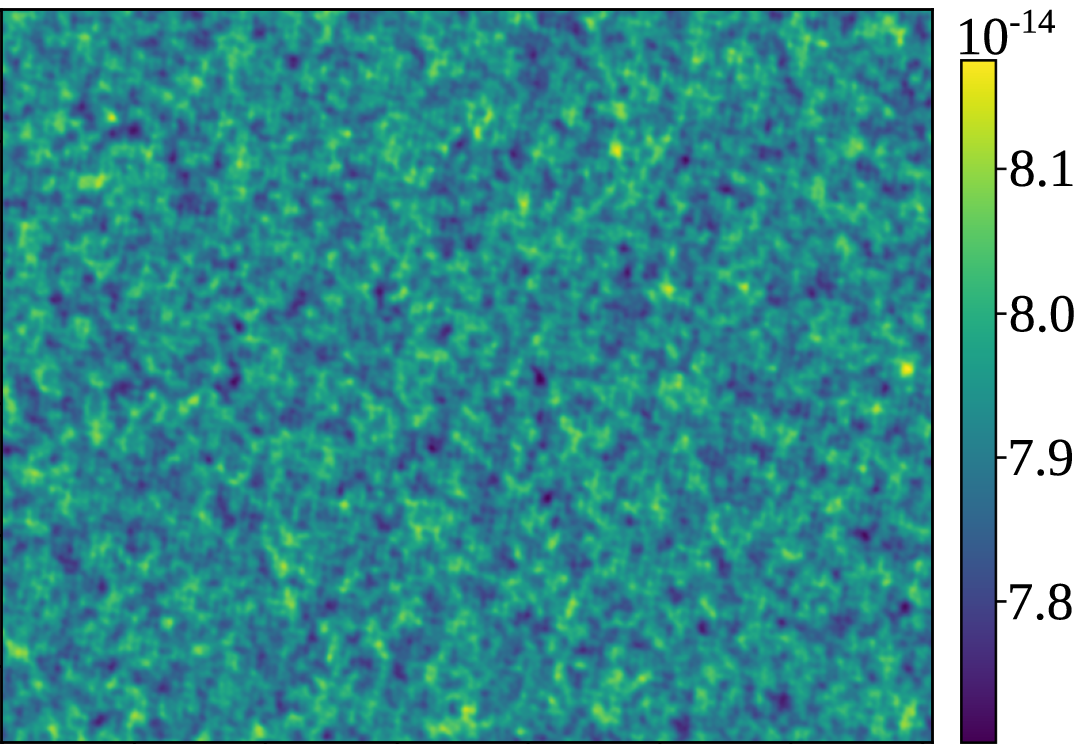}
        \caption{$l_\mathrm{c}=10^{-2}$.}
    \end{subfigure}\\
        \begin{subfigure}{0.32\textwidth}
        \includegraphics[height=0.17\textheight]{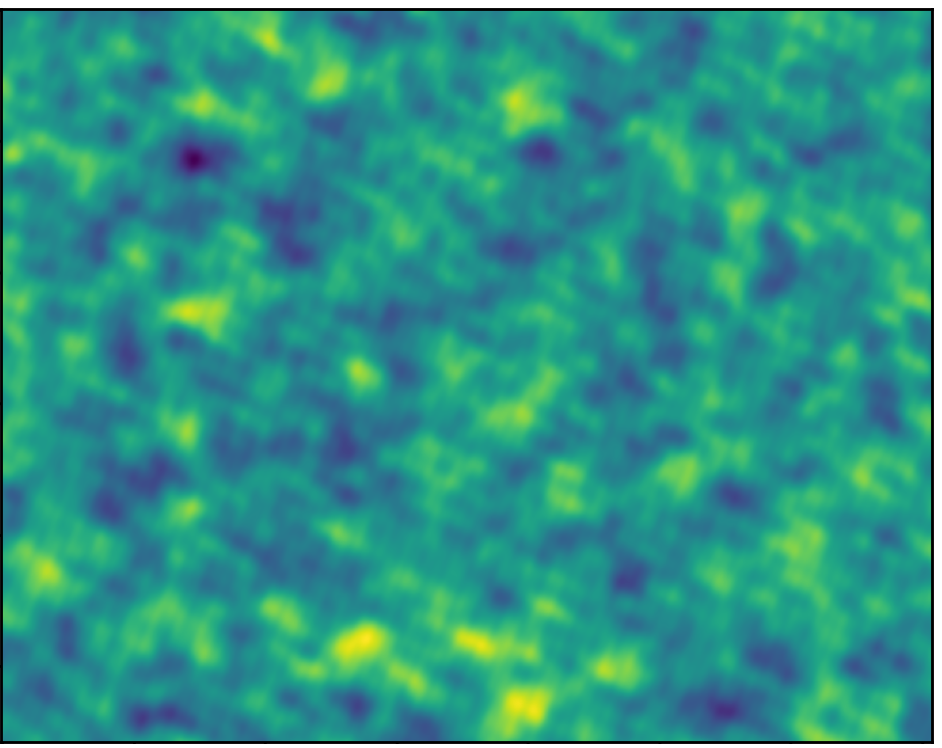}
        \caption{$l_\mathrm{c}=3\cdot10^{-2}$.}
    \end{subfigure}
    \begin{subfigure}{0.32\textwidth}
        \includegraphics[height=0.17\textheight]{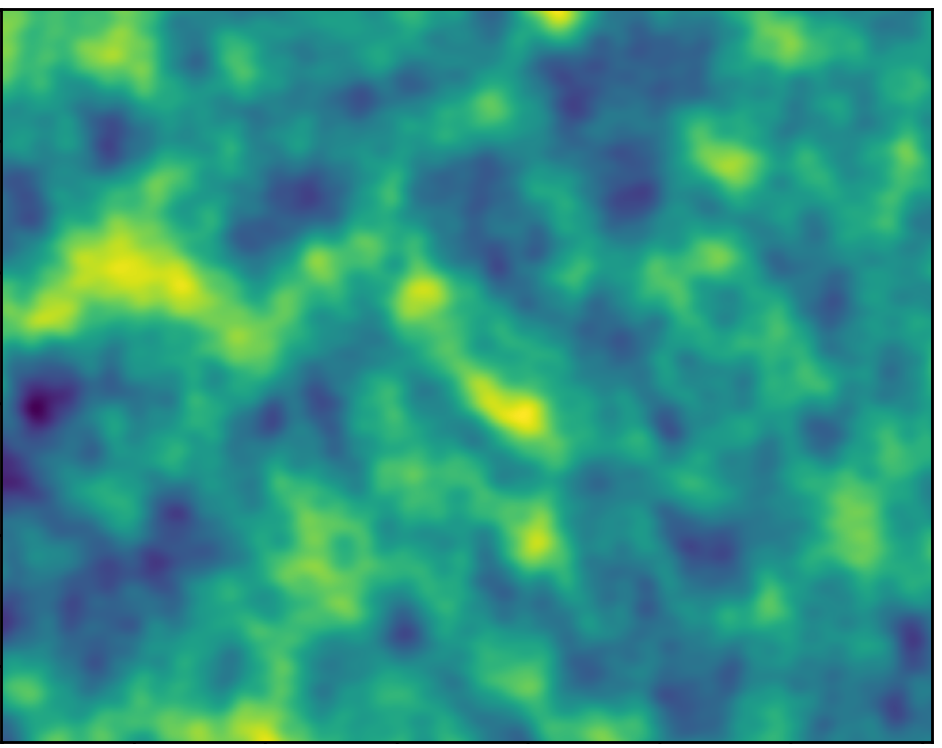}
        \caption{$l_\mathrm{c}=5\cdot10^{-2}$.}
    \end{subfigure}
    \begin{subfigure}{0.32\textwidth}
        \includegraphics[height=0.17\textheight]{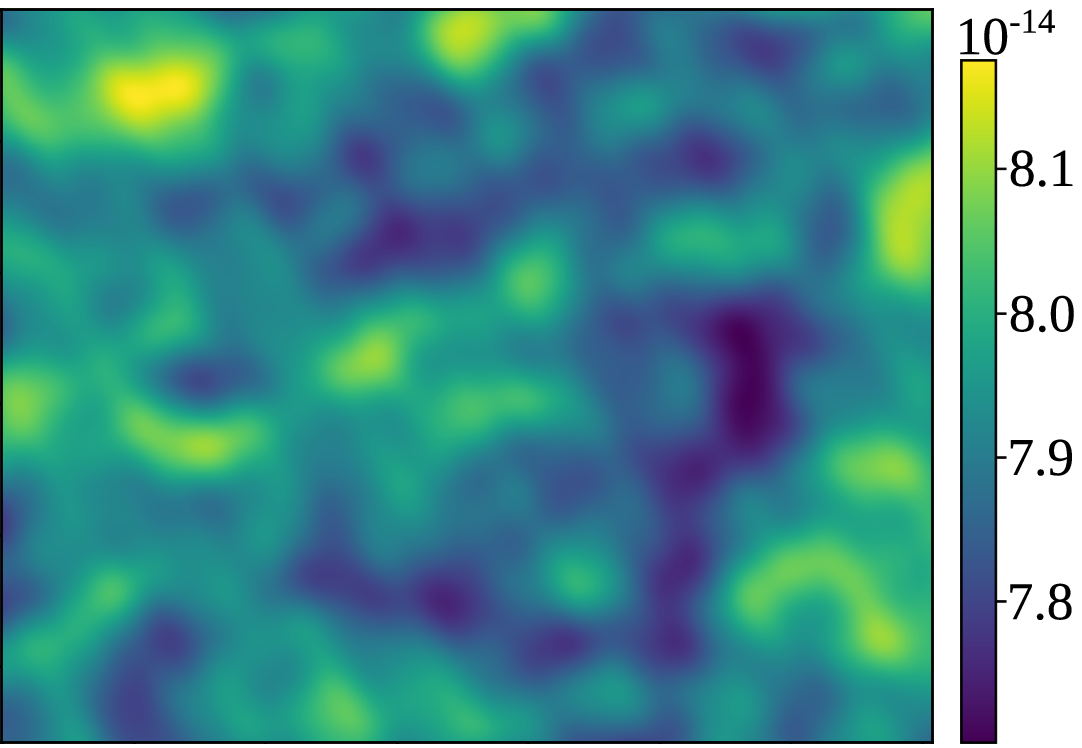}
        \caption{$l_\mathrm{c}=7.5\cdot10^{-2}$.}
    \end{subfigure}\\
        \begin{subfigure}{0.32\textwidth}
        \includegraphics[height=0.17\textheight]{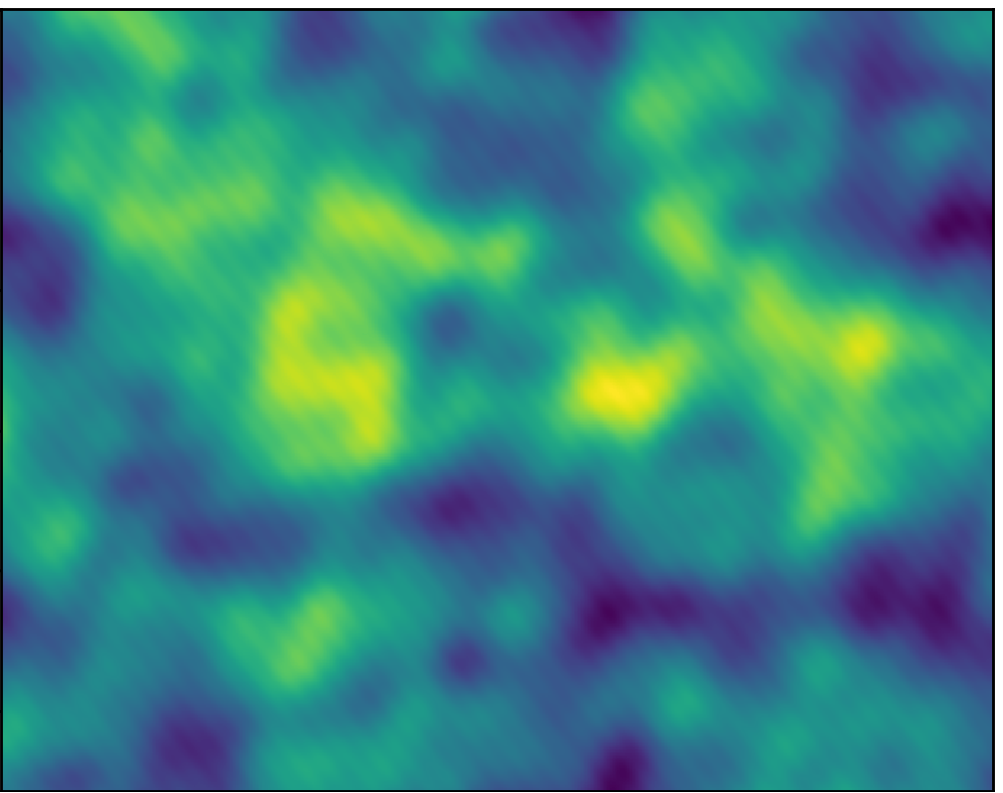}
        \caption{$l_\mathrm{c}=9\cdot10^{-2}$.}
    \end{subfigure}
    \begin{subfigure}{0.32\textwidth}
        \includegraphics[height=0.17\textheight]{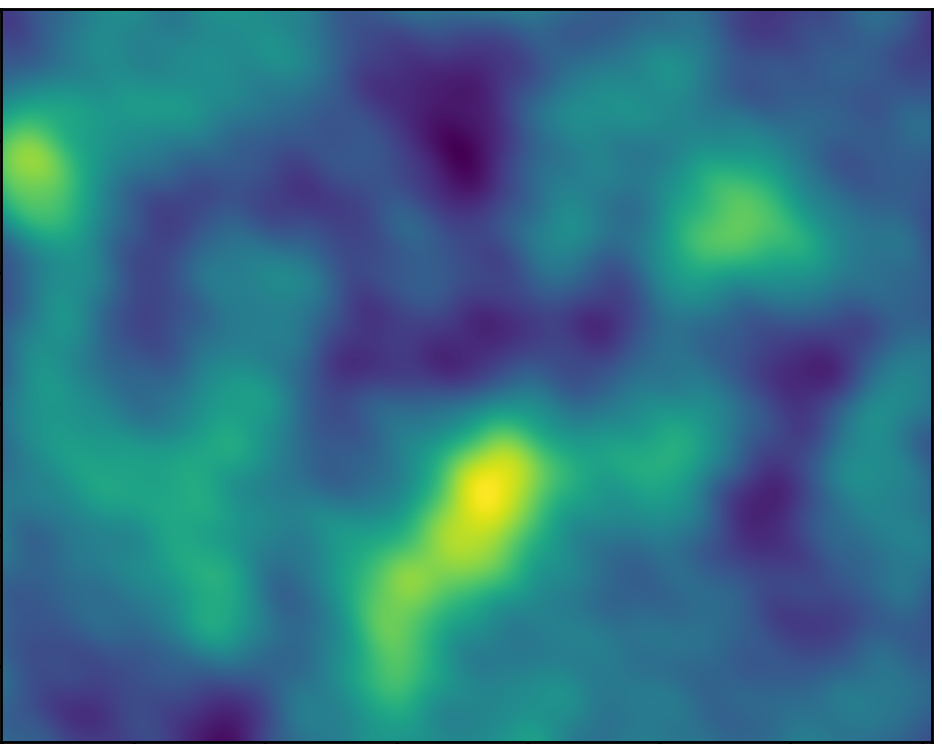}
        \caption{$l_\mathrm{c}=0.1$.}
    \end{subfigure}
    \begin{subfigure}{0.32\textwidth}
        \includegraphics[height=0.17\textheight]{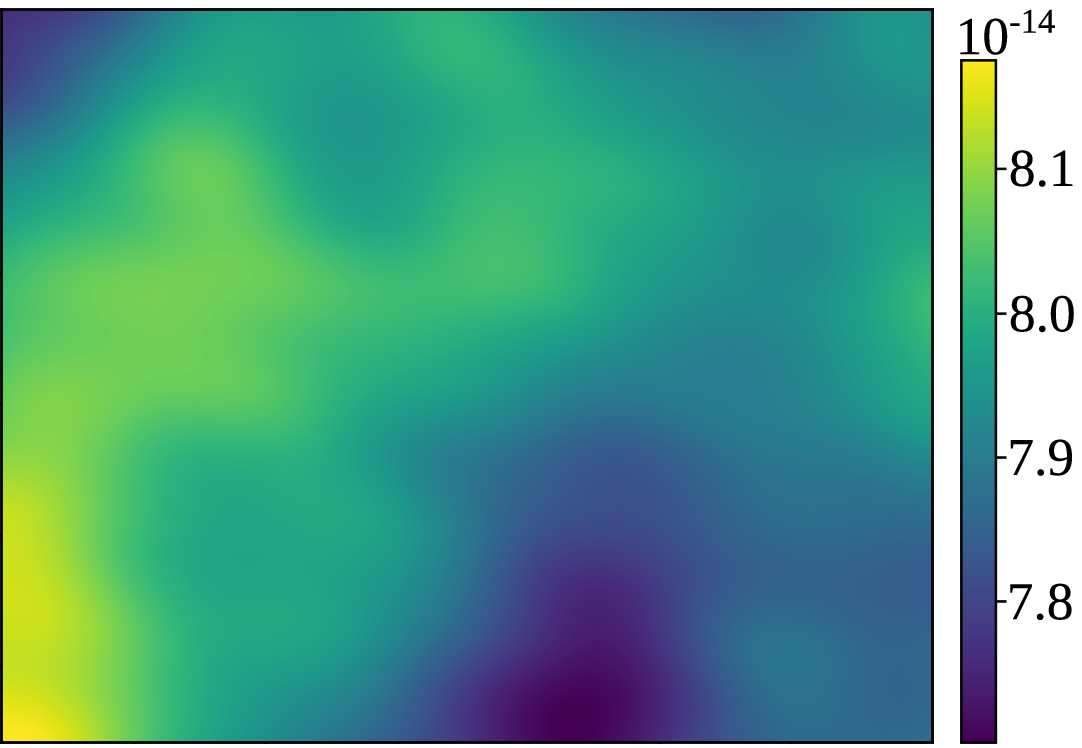}
        \caption{$l_\mathrm{c}=0.25$.}
    \end{subfigure}\\
    \caption{The examples of the permeability maps $k(x,y)$ for different correlation lengths used as input parameters for water-polymer flooding simulations. The  root-mean-square deviation $\delta_\mathrm{d}$ is fixed at $\pm 0.03 k_0$. The maps provided correspond to fine-, medium-, and coarse-grained formation structures.}
    \label{fig:perm_lc}
\end{figure}

The coarseness of the permeability structure was chosen as a stochastic parameter of reservoir properties. Maps with random permeability were generated using the GPFlowSampling package\,\cite{bib:gpflow} for building Gaussian field models, which reduces the sampling complexity to being linear in the length of the vector we condition on. Such a piece of software allows us to vary the permeability parameters in the most appropriate way. Besides the normal shape of the distribution and the root-mean-square deviation $\delta_\mathrm{d}$ from the mean $k_0$, the magnitude of the correlation of permeability values between different spatial coordinates $l_\mathrm{c}$ is of particular interest. This parameter characterizes the distribution of the entire process as a whole, not the Gaussian values at individual points. 
The extent of correlation affects how fast the dependencies between different spatial coordinates decay as the distance between them increases. For example, when $l_\mathrm{c}$ is close to zero, the trajectories resemble noise; when it is very large, they resemble constant functions. This mathematical approximation makes it possible to model the coarseness of the reservoir structure without changing the parameters or the shape of the Gaussian distribution of permeability values. Numerical calculations were performed for a range of correlation values from  $l_\mathrm{c}$=5$\cdot 10^{-4}$ to $l_\mathrm{c}$=2.5$\cdot 10^{-1}$, corresponding to fine-, medium-, and coarse-grained formation structures. Note that the classification of reservoir structure into fine-, medium-, and coarse-grained formations is informal in nature and based on visualization of permeability maps. The typical examples of the $k(x,y)$ are presented in Fig.\,\ref{fig:perm_lc}. It is worth recalling here that all correlation lengths are normalized to the reservoir dimension $L$.

\subsection{DuMu\textsuperscript{x}}
DuMu$^\mathrm{x}$~\cite{bib:dumux11,bib:dumux20} is a framework for porous media flow based on the DUNE (Distributed and Unified Numerics Environment) toolbox~\cite{ans-DUNE}. Both DuMu$^\mathrm{x}$ and DUNE are open-source software and extensively use C++ template metaprogramming. While DUNE provides generic interfaces for implementing various discretization schemes, DuMu$\mathrm{x}$ uses the finite volume method exclusively. For simulation with DuMu$^\mathrm{x}$, a fully implicit scheme was used in this work (version 3 of the package gives the most support for fully implicit schemes). As for the flux approximation, DuMu$^\mathrm{x}$ employs the cell-centered scheme (for both two-point and multi-point flux approximation flavors) and the vertex-centered scheme, also known as the box method. The latter was involved in this work. Finally, of the many linear solvers that DUNE contains, the implementation of the biconjugate gradient stabilized method with an algebraic multigrid preconditioner (AMGBiCGSTAB) was used. DuMu$^\mathrm{x}$ accepts grids in a variety of formats, including the well-known .msh format~\cite{gmsh}.

\subsection{Modelling parameters}
Numerical experiments were carried out on a rectangular grid of 0.4 million elements with a grid cell size of 5.5\,cm. The dimensions of the computational domain were $W\times L=31.14\,$m\,$\times \,40\,$m. The constant-injection rate condition $q_0=10$\,m$^2$/d (injection well) was set at the left boundary of the area and the zero pressure value (production well) was set at the right side. 
We examine the scenario of water injection into a reservoir completely filled with polymer at a concentration of $c_\mathrm{max}=0.0015$. A quadratic function of viscosity was considered to optimize and speed up the numerical calculation process while reproducing near-to-real conditions. This simplification does not affect the estimation of the growth rate of the viscous fingers and is equivalent to considering the velocity of the fingers in the case of a polymer slug\,\cite{Linear}. At $c_\mathrm{max}$ polymer concentration, the relative viscosity increases by a factor of $M=20$. The following parameters were used: mean permeability $k_0=80$\,mD, water viscosity $\mu_\mathrm{w}=0.3$\,cP, and porosity $\phi=0.188$. Hence, as we simulate a two-dimensional problem, the surface area of the reservoir and the porosity value determine the total pore volume 1PV=$\phi LW = 234.2$~m$^2$.

The evolution of the boundary of the water-polymer mixing zone over time was analyzed (see Figure\,\ref{fig:slug}). Due to the variability of the spatial distribution of the permeability maps, $k(x,y)$, the velocities of the anterior and posterior boundaries of the mixing zone turn out to be different from each other in each simulation. In what follows, we will be interested in the value of $x_\mathrm{f}$ corresponding to the maximum distance between the injection well and the front boundary of the mixing zone (the distance to the tip of the fastest finger). This boundary was defined as the point farthest from the injection well where the concentration of the polymer is above 0.95$c_\mathrm{max}$. In the analysis of the output distributions, the polymer concentrations for the data located near the lateral boundaries of the reservoir around 0.05$W$ were not considered, --- to exclude the influence of the boundary conditions in this region on the final statistics of the growth rate.

\begin{table}[H]
 \centering
 \begin{tabular}{c|l}
 \hline\hline
Reservoir parameters  & Data values\\
 \hline
Initial concentration of polymer, $c_\mathrm{max}$ & 0.0015\\
Mean permeability, $k_0$  & 80 [mD]\\
Viscosity of water, $\mu_\mathrm{w}$  & 0.3 [cP]\\
Porosity, $\phi$  & 0.188\\
Surface injection rate, $q_0$ & 10 [m$^2$/d]\\
Viscosity ratio parameter, $M$ & 20\\
Computational domain length, $L$ & 40 [m]\\
Computational domain width, $W$ & 31.4 [m]\\
Quadratic viscosity function, $\mu(c)$ & $\mu_\mathrm{w}+ \mu_\mathrm{w}(M-1)(c/c_\mathrm{max})^2$ [cP]\\
\hline
 \end{tabular}
 \caption{Simulation parameters used to build a stochastic model for the velocity of viscous fingers in a water-polymer system.}
 \label{tab:params}
\end{table}

A complete list of parameters involved in the implementation of numerical calculations is given in Table\,\ref{tab:params}. The dependencies of the motion of the point $x_\mathrm{f}$ (see Figure\,\ref{fig:slug}) on the injection time were constructed, as reported in\,\cite{Linear}, to determine the growth rate of viscous fingers through the polymer. The presence of variability in the permeability profile causes significant scatter in the results, which depends on the degree of coarseness, or, in other words, on the correlation length $l_\mathrm{c}$. 

\section{Influence of the correlation length of reservoir permeability}
\label{sec:cl}

\begin{figure}[t!]
    \centering
    \begin{subfigure}{0.32\textwidth}
        \includegraphics[height=0.17\textheight]{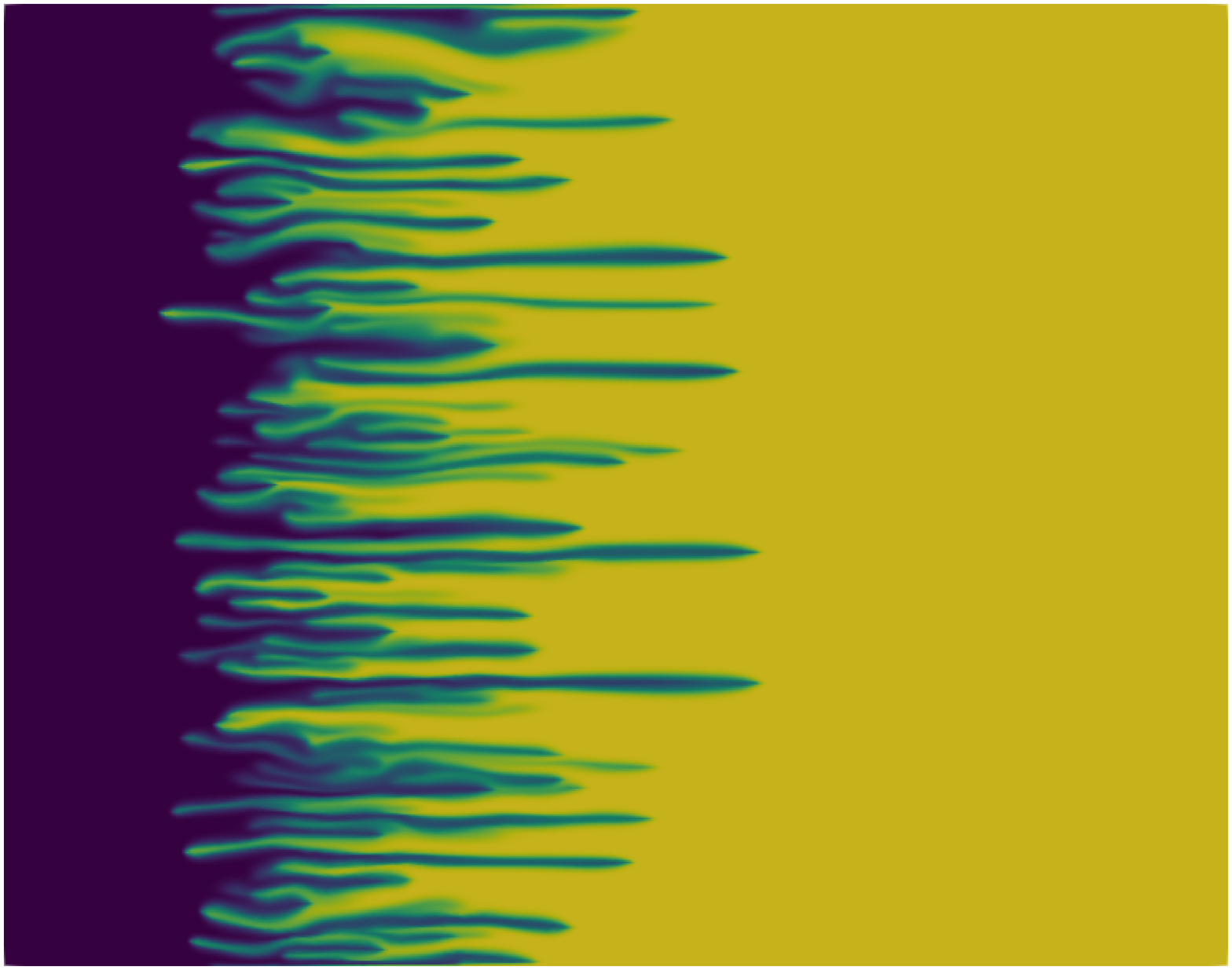}
        \caption{$l_\mathrm{c}=10^{-4}$.}
    \end{subfigure}
    \begin{subfigure}{0.32\textwidth}
        \includegraphics[height=0.17\textheight]{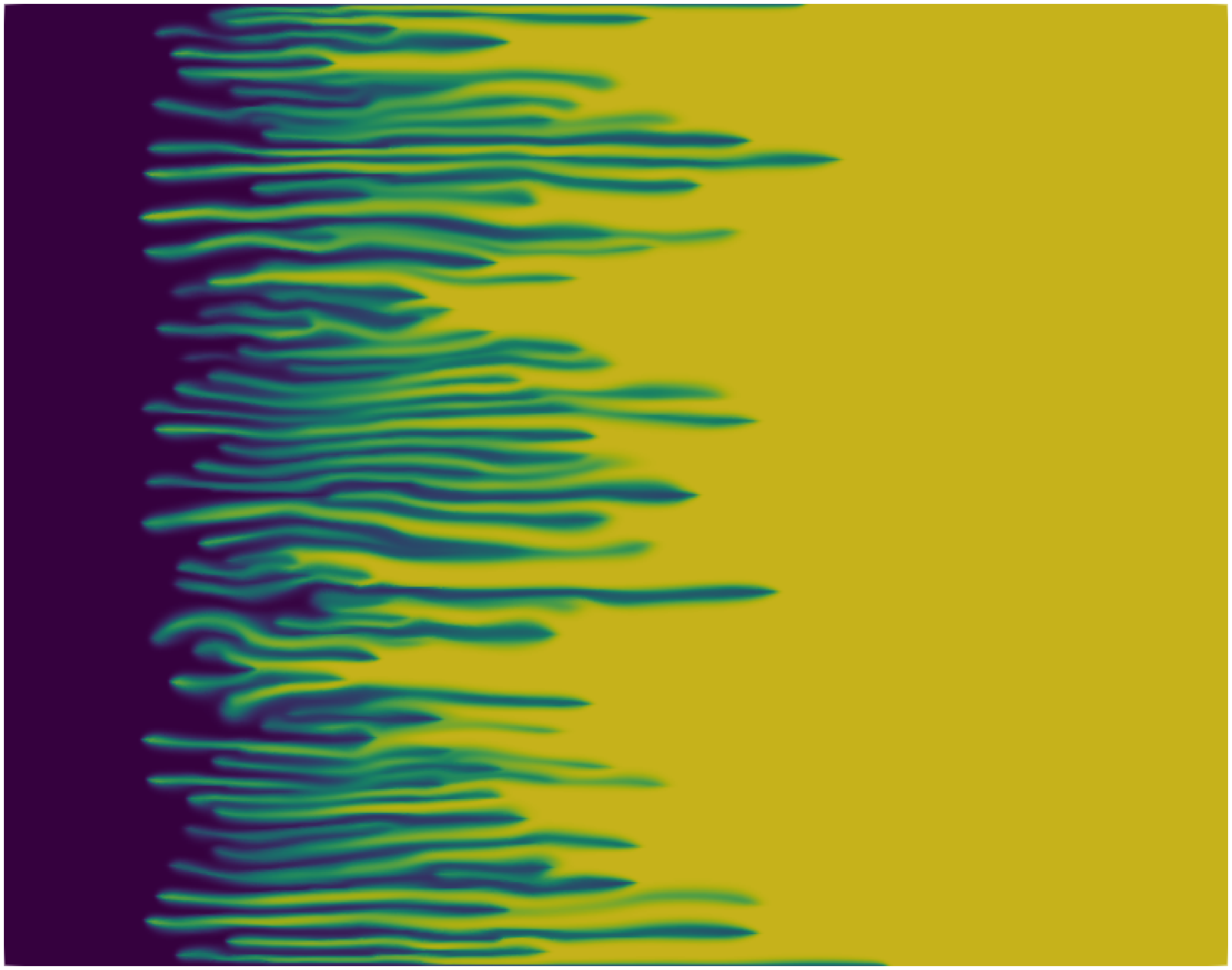}
        \caption{$l_\mathrm{c}=5\cdot 10^{-3}$.}
    \end{subfigure}
    \begin{subfigure}{0.32\textwidth}
        \includegraphics[height=0.17\textheight]{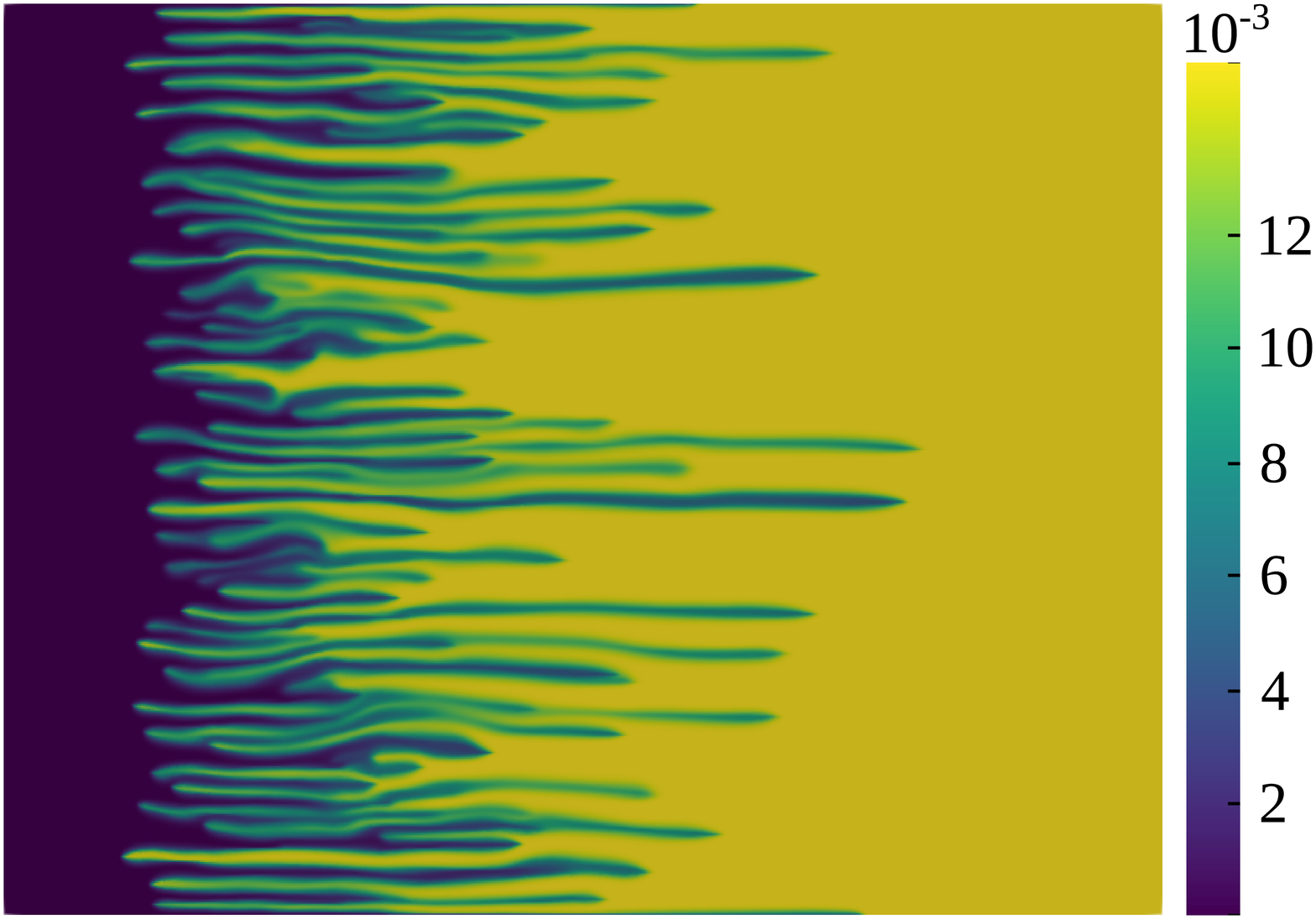}
        \caption{$l_\mathrm{c}=10^{-2}$.}
    \end{subfigure}\\
        \begin{subfigure}{0.32\textwidth}
        \includegraphics[height=0.17\textheight]{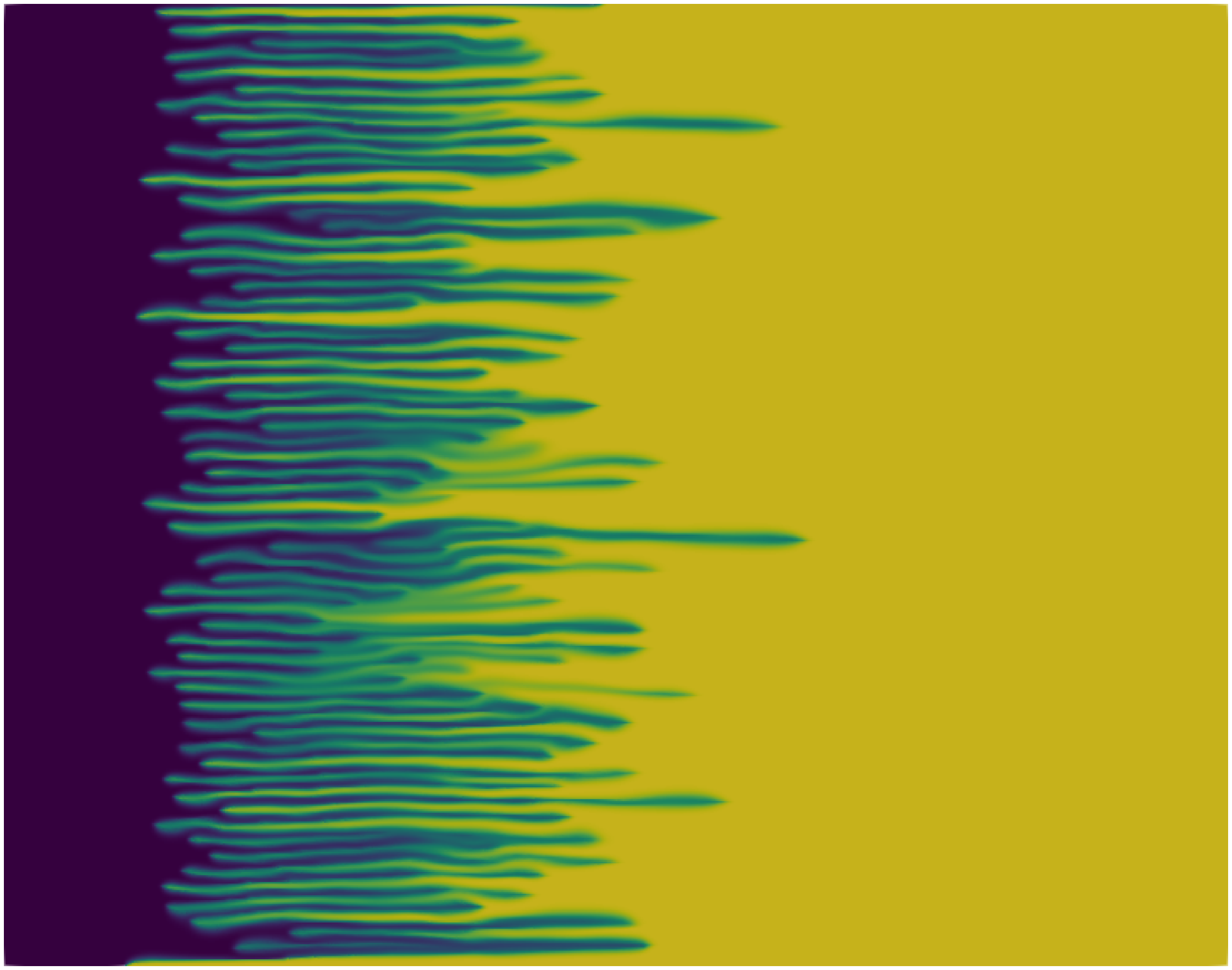}
        \caption{$l_\mathrm{c}=3\cdot 10^{-2}$.}
    \end{subfigure}
    \begin{subfigure}{0.32\textwidth}
        \includegraphics[height=0.17\textheight]{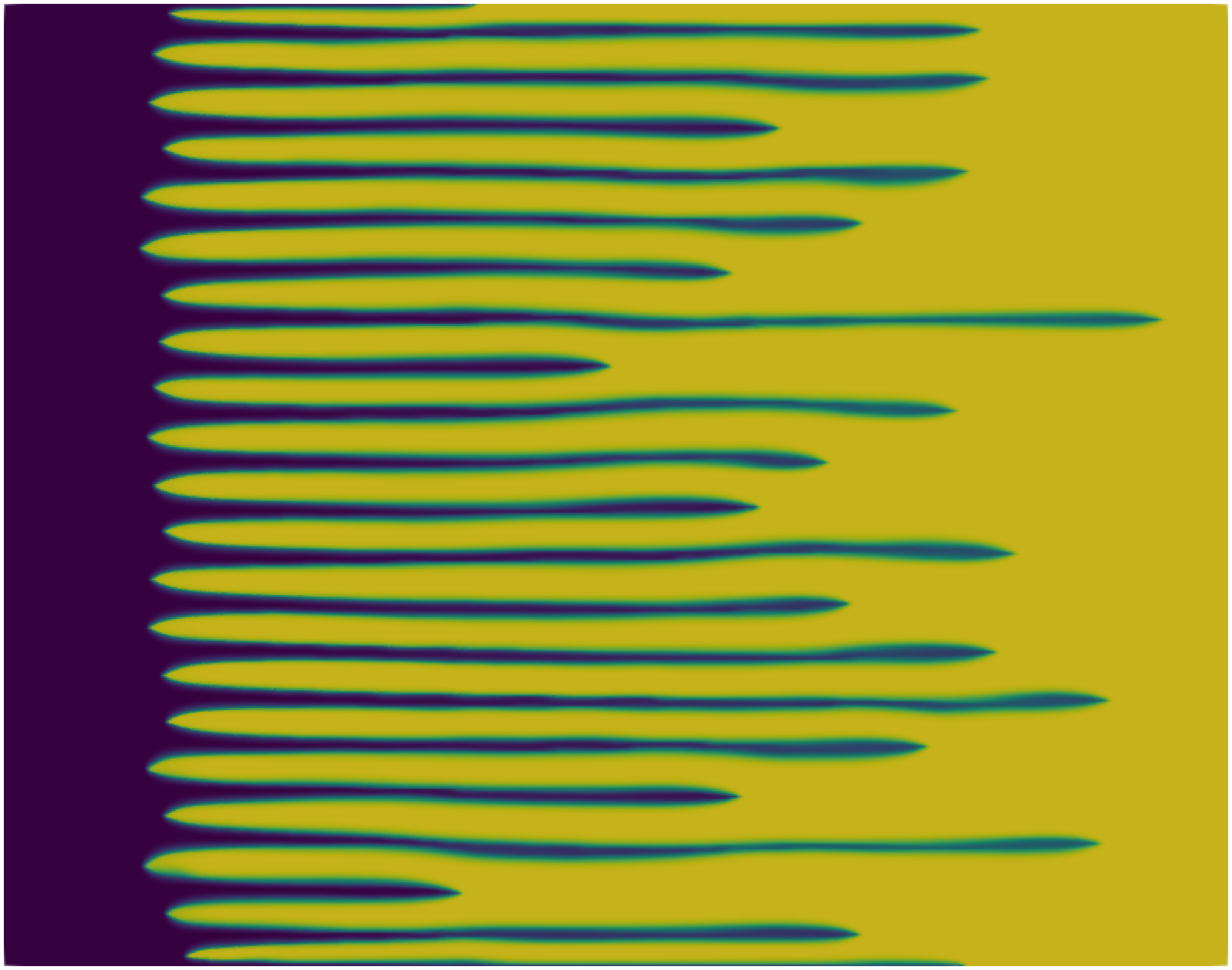}
        \caption{$l_\mathrm{c}=5\cdot 10^{-2}$.}
    \end{subfigure}
    \begin{subfigure}{0.32\textwidth}
        \includegraphics[height=0.17\textheight]{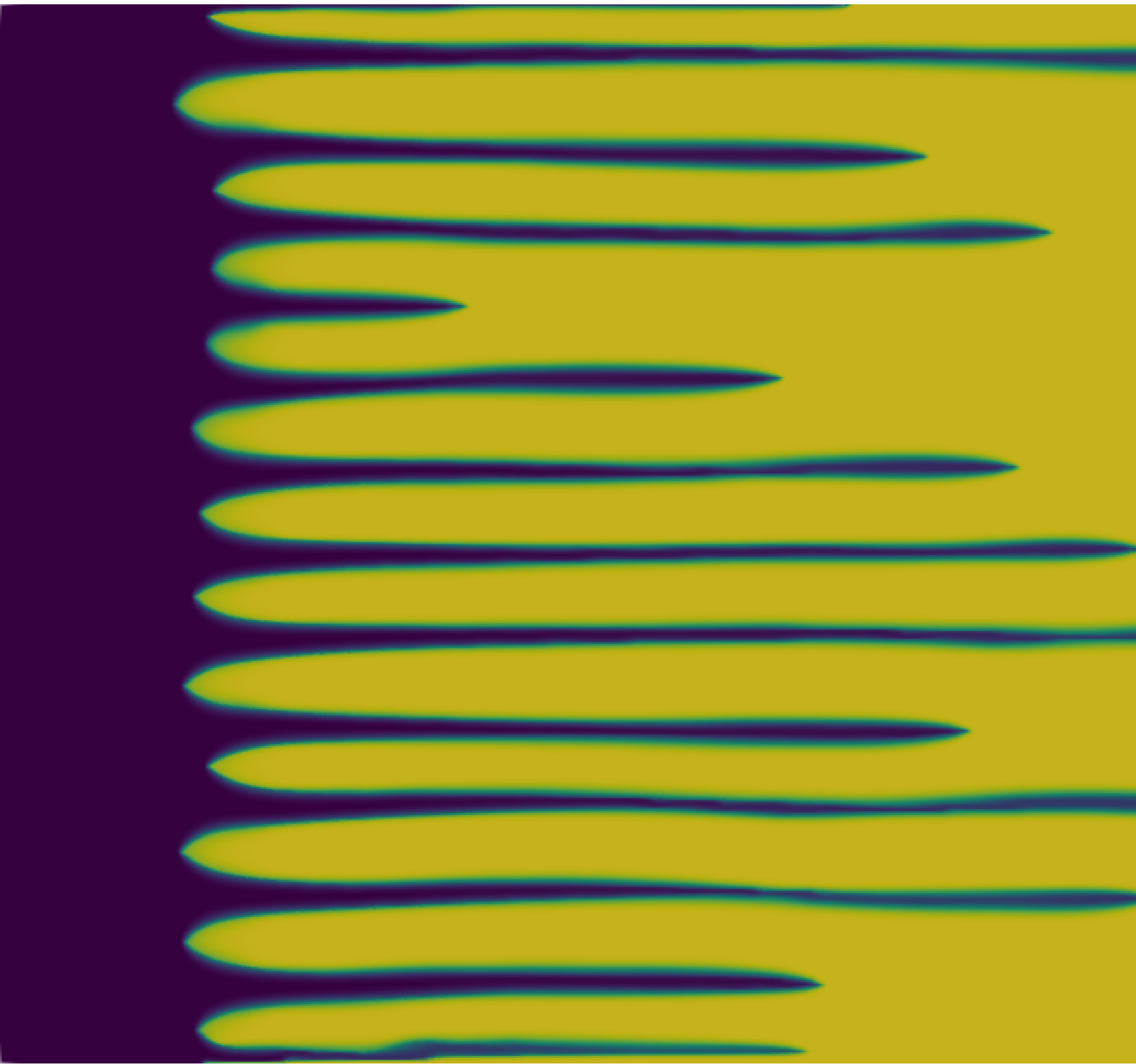}
        \caption{$l_\mathrm{c}=7.5\cdot 10^{-2}$.}
    \end{subfigure}\\
        \begin{subfigure}{0.32\textwidth}
        \includegraphics[height=0.17\textheight]{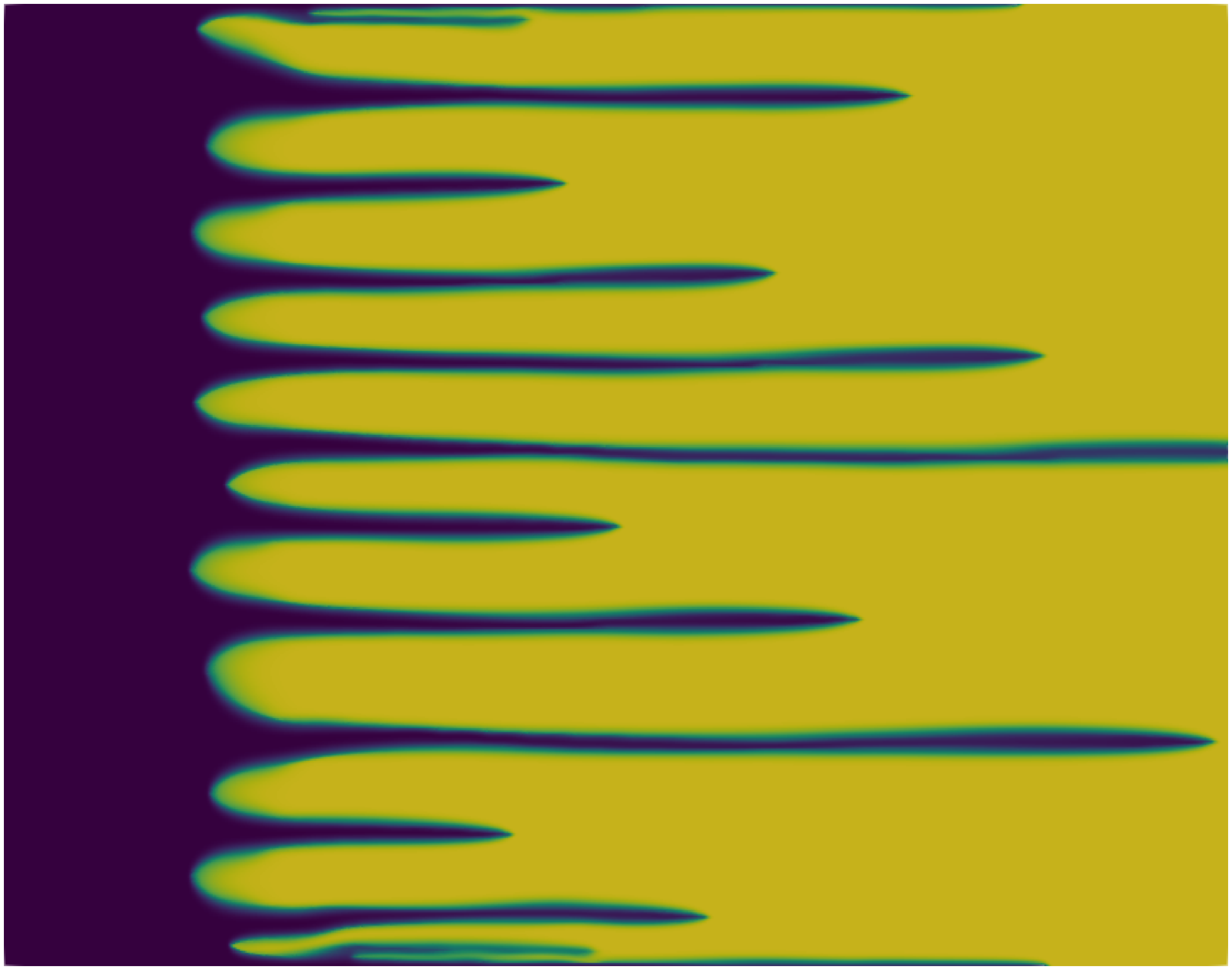}
        \caption{$l_\mathrm{c}=9\cdot 10^{-2}$.}
    \end{subfigure}
    \begin{subfigure}{0.32\textwidth}
        \includegraphics[height=0.17\textheight]{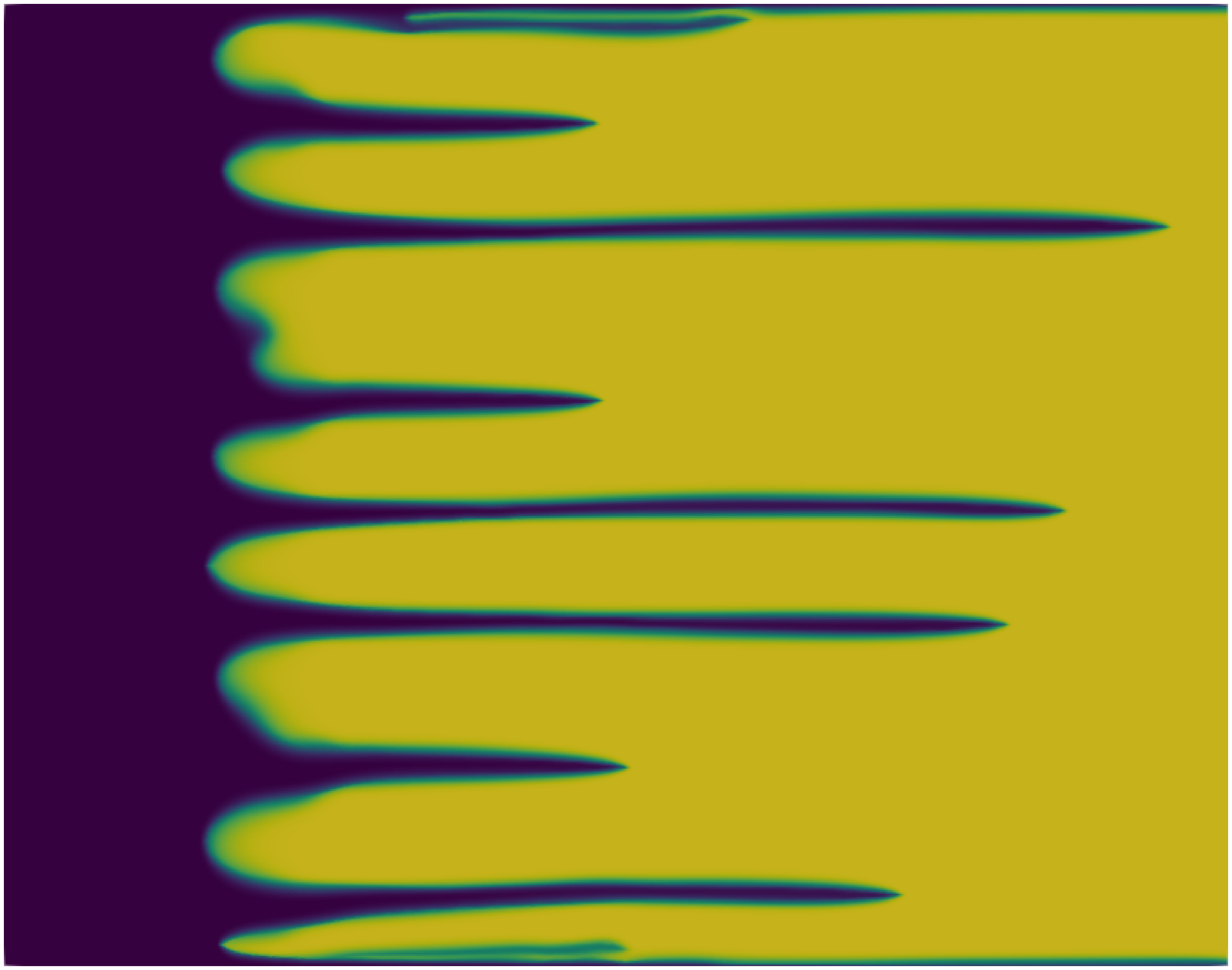}
        \caption{$l_\mathrm{c}=0.1$.}
    \end{subfigure}
    \begin{subfigure}{0.32\textwidth}
        \includegraphics[height=0.17\textheight]{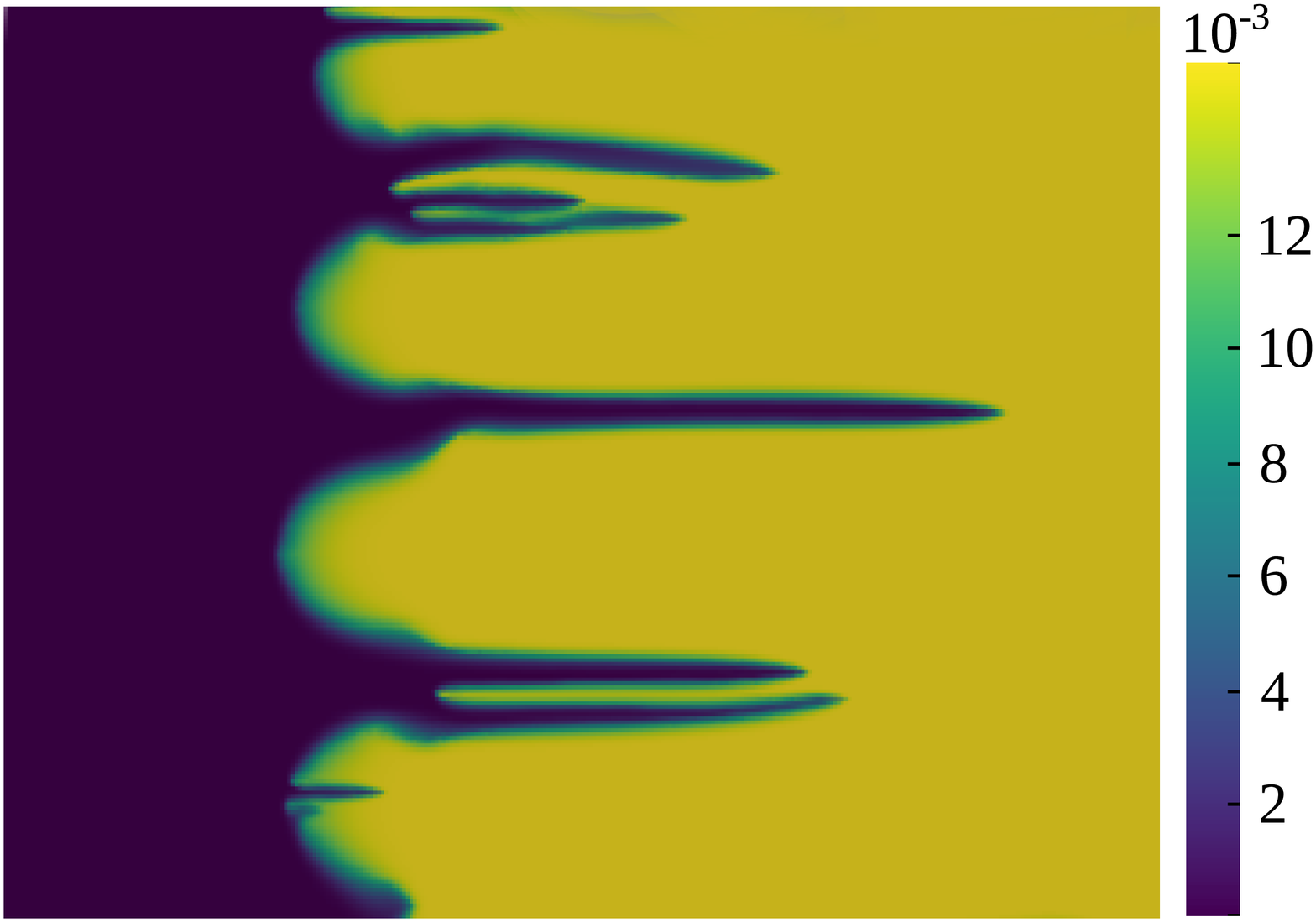}
        \caption{$l_\mathrm{c}=0.25$.}
    \end{subfigure}\\
    \caption{The viscous fingers in the polymer bank formed after 0.4\,PV water injection. The distributions were obtained for permeability maps from Figure\,\ref{fig:perm_lc} as input parameters for numerical simulations.}
    \label{fig:fingers_lc}
\end{figure}

\begin{figure}[t!]
    \centering
    \begin{subfigure}{0.3\textwidth}
        \includegraphics[height=0.15\textheight]{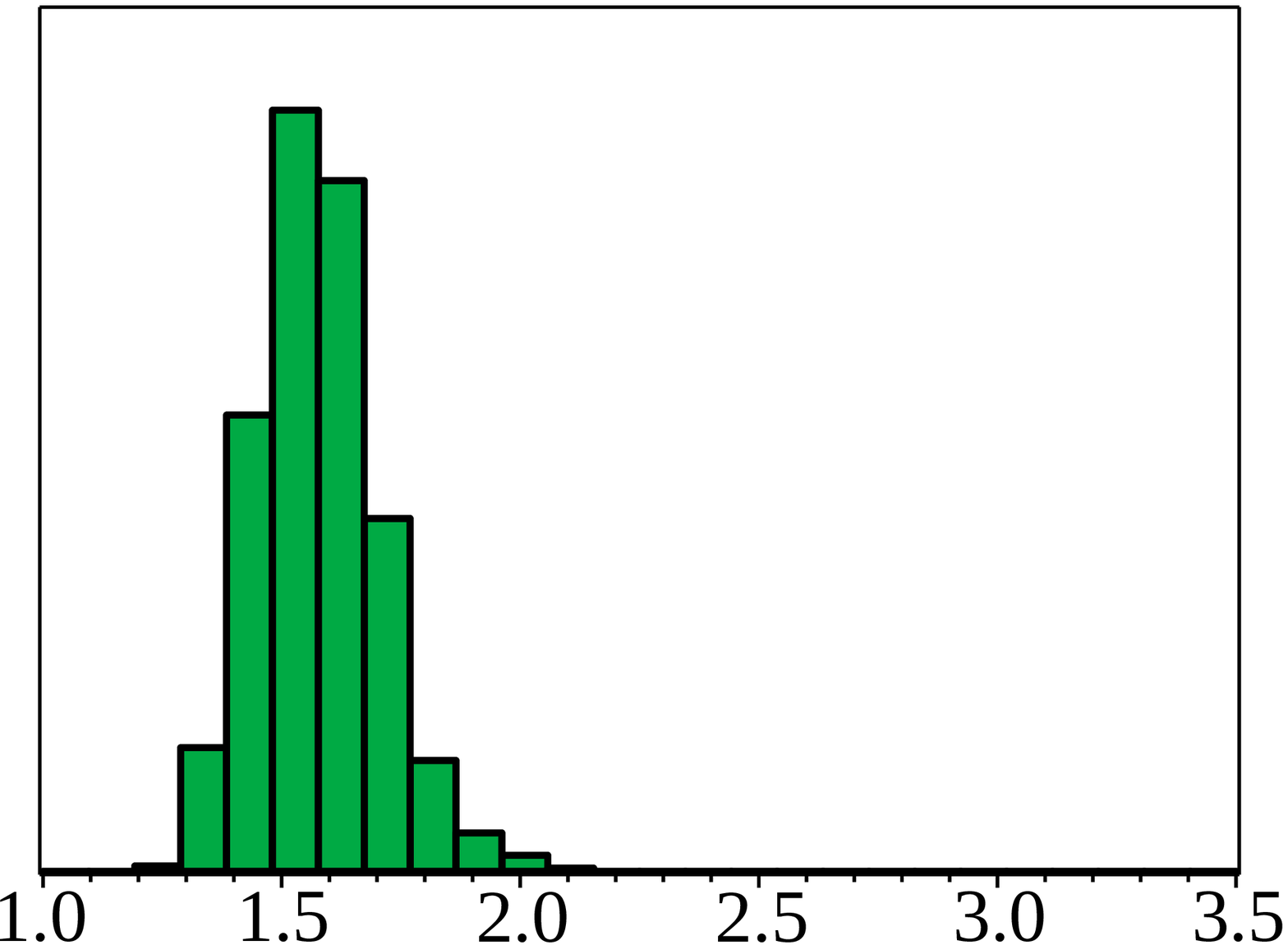}
        \caption{$l_\mathrm{c}=10^{-4}$.}
    \end{subfigure}
    \begin{subfigure}{0.3\textwidth}
        \includegraphics[height=0.15\textheight]{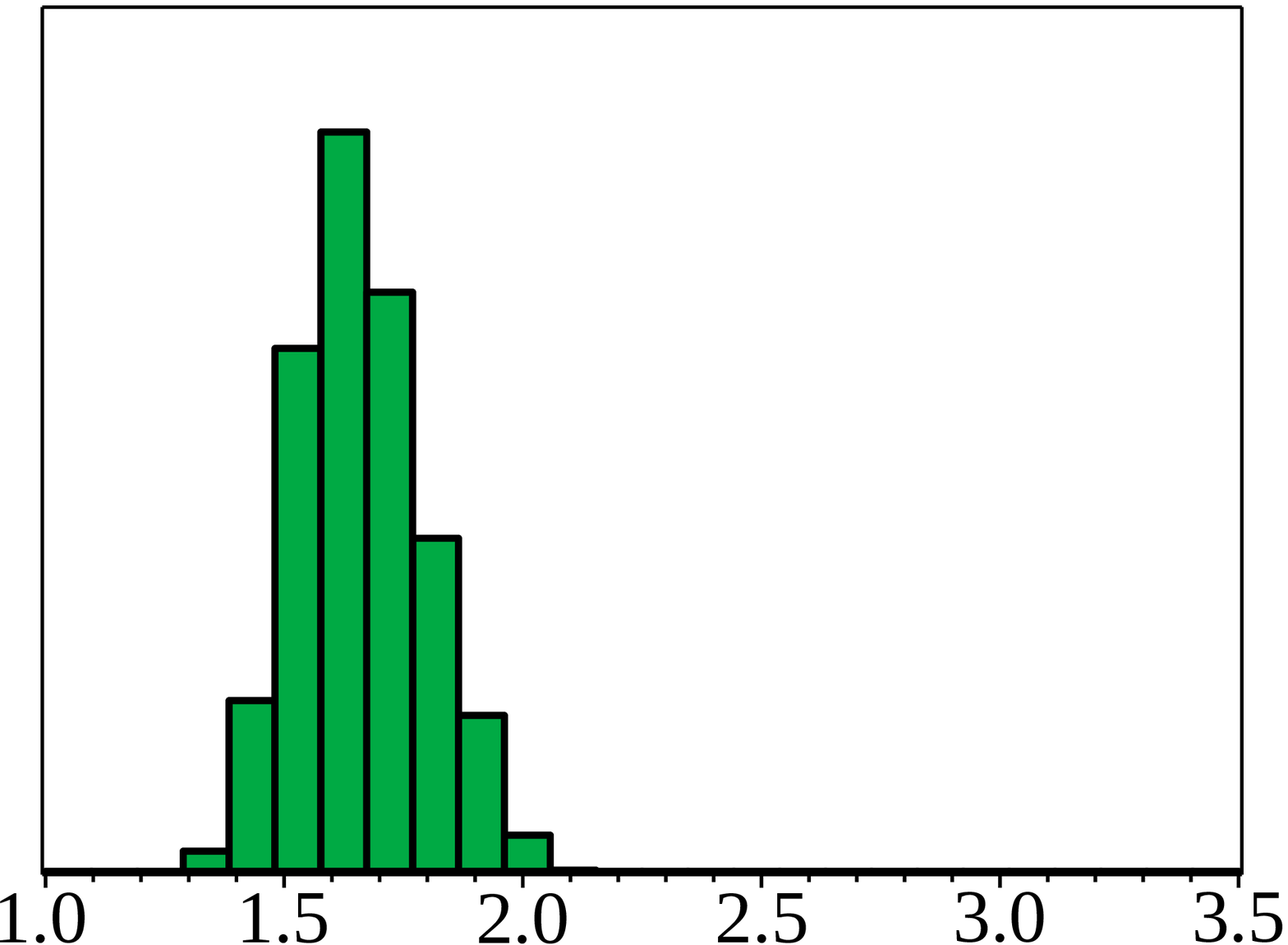}
        \caption{$l_\mathrm{c}=5\cdot 10^{-3}$.}
    \end{subfigure}
    \begin{subfigure}{0.3\textwidth}
        \includegraphics[height=0.15\textheight]{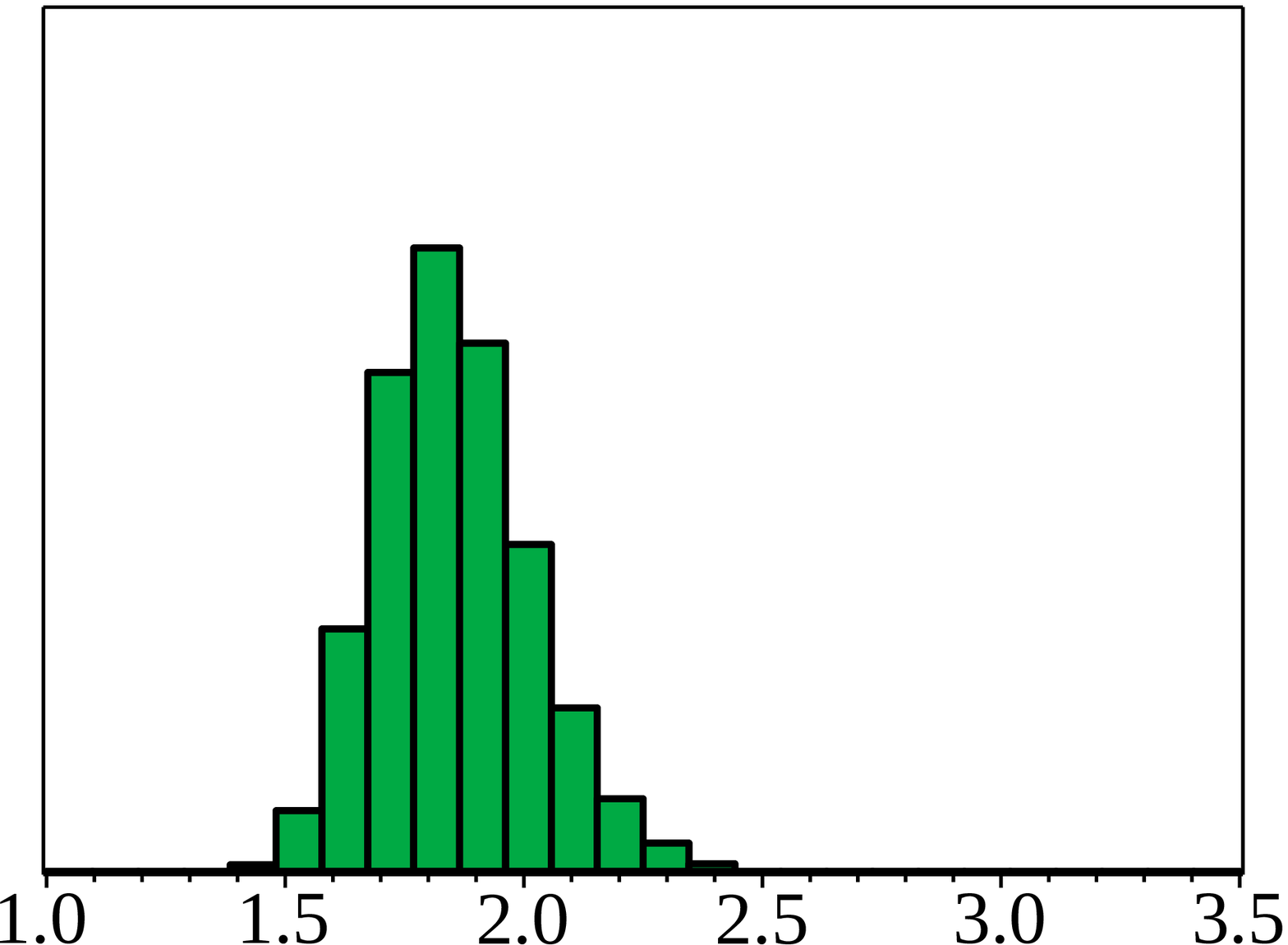}
        \caption{$l_\mathrm{c}=10^{-2}$.}
    \end{subfigure}\\
        \begin{subfigure}{0.3\textwidth}
        \includegraphics[height=0.15\textheight]{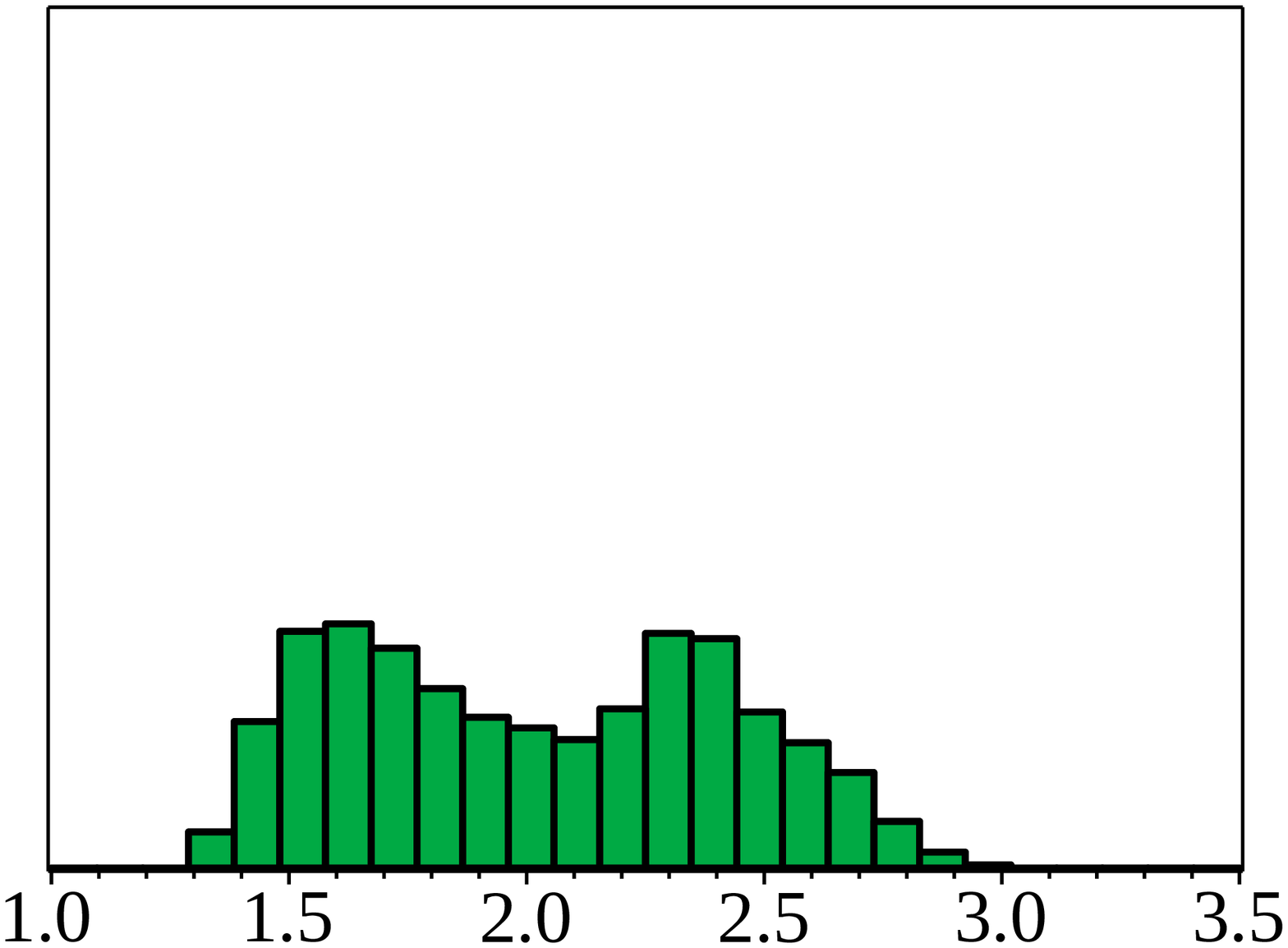}
        \caption{$l_\mathrm{c}=3\cdot 10^{-2}$.}
    \end{subfigure}
    \begin{subfigure}{0.3\textwidth}
        \includegraphics[height=0.15\textheight]{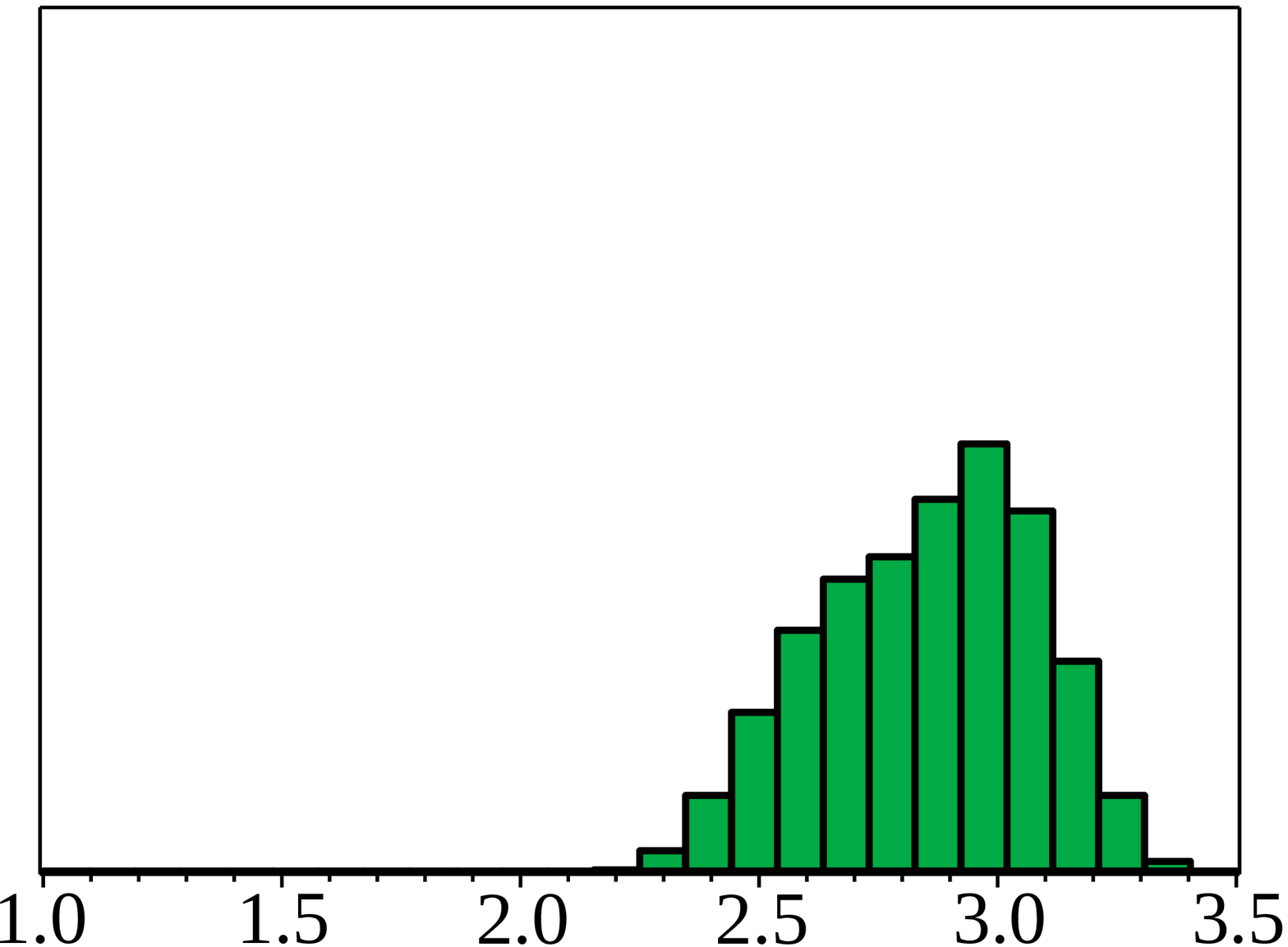}
        \caption{$l_\mathrm{c}=5\cdot 10^{-2}$.}
    \end{subfigure}
    \begin{subfigure}{0.3\textwidth}
        \includegraphics[height=0.15\textheight]{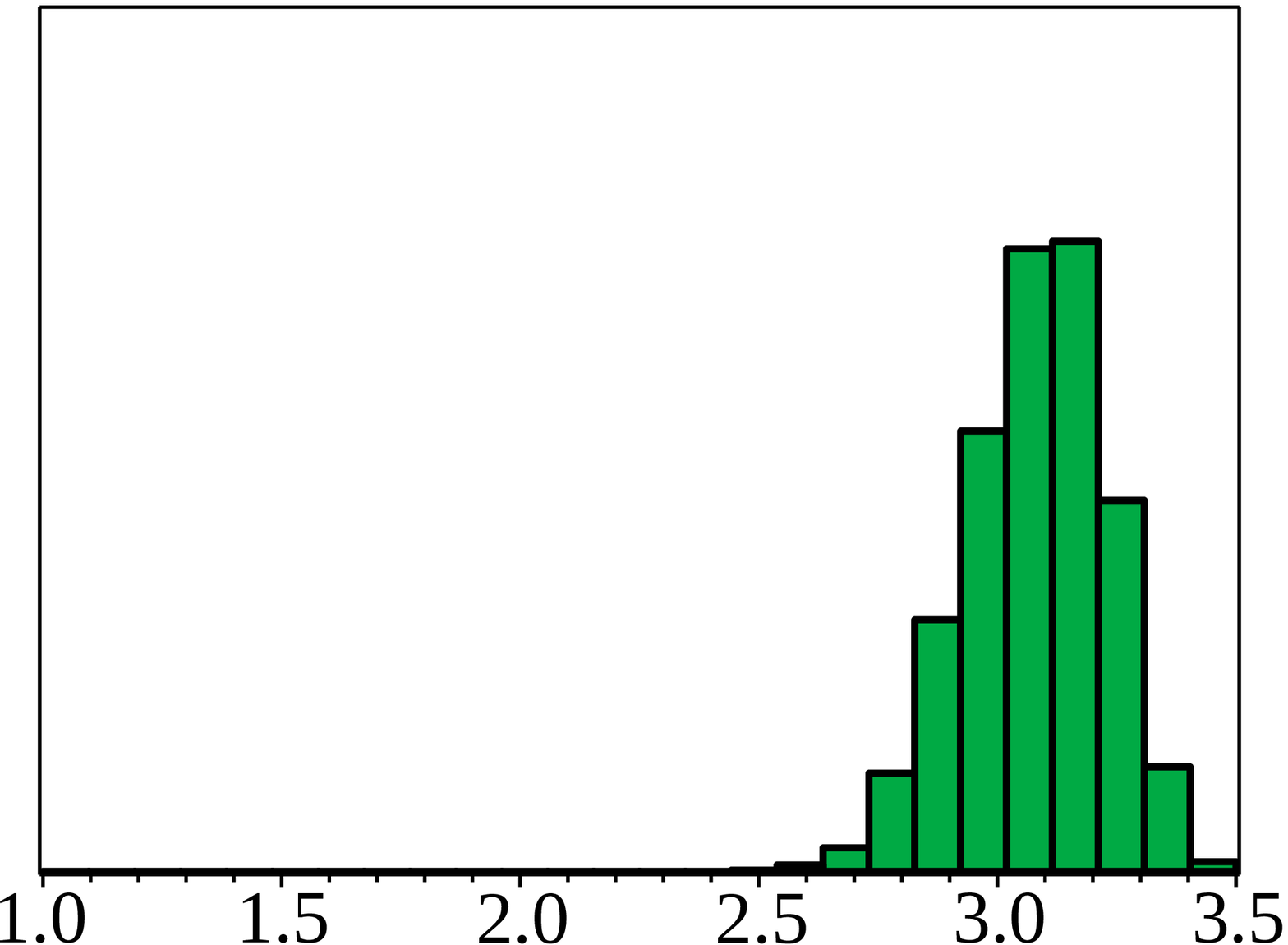}
        \caption{$l_\mathrm{c}=7.5\cdot 10^{-2}$.}
    \end{subfigure}\\
        \begin{subfigure}{0.3\textwidth}
        \includegraphics[height=0.15\textheight]{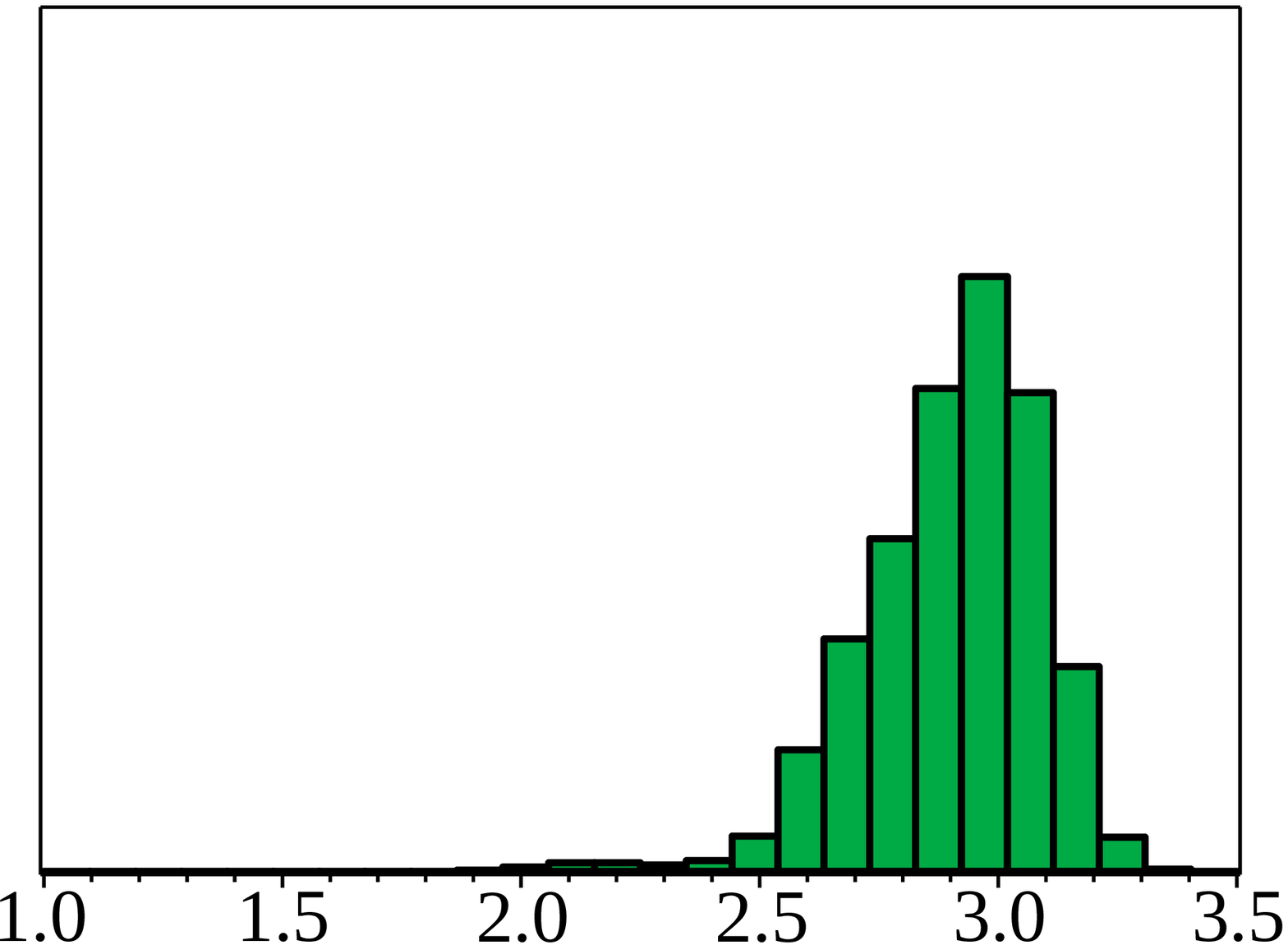}
        \caption{$l_\mathrm{c}=9\cdot 10^{-2}$.}
    \end{subfigure}
    \begin{subfigure}{0.3\textwidth}
        \includegraphics[height=0.15\textheight]{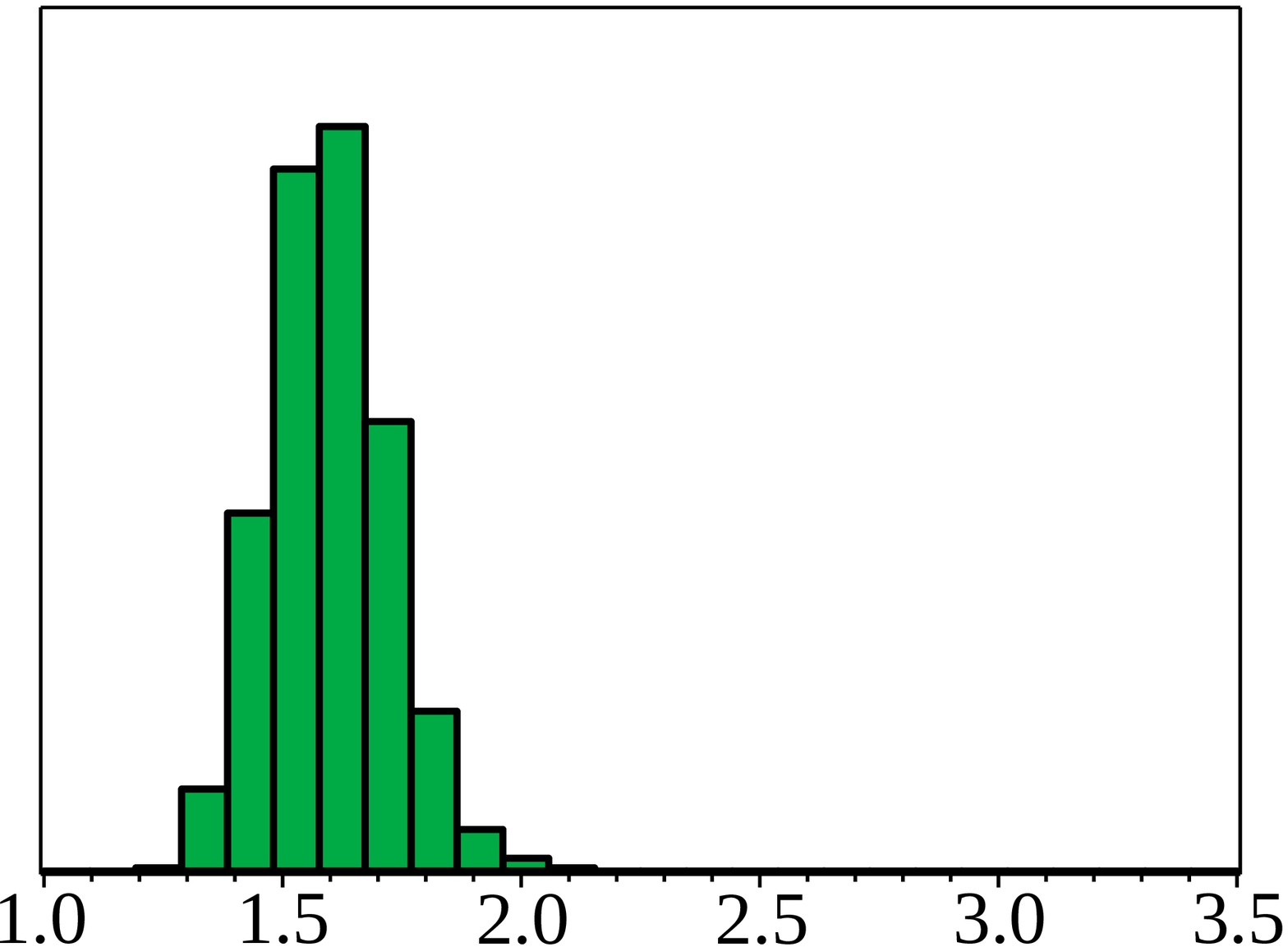}
        \caption{$l_\mathrm{c}=0.1$.}
    \end{subfigure}
    \begin{subfigure}{0.3\textwidth}
        \includegraphics[height=0.15\textheight]{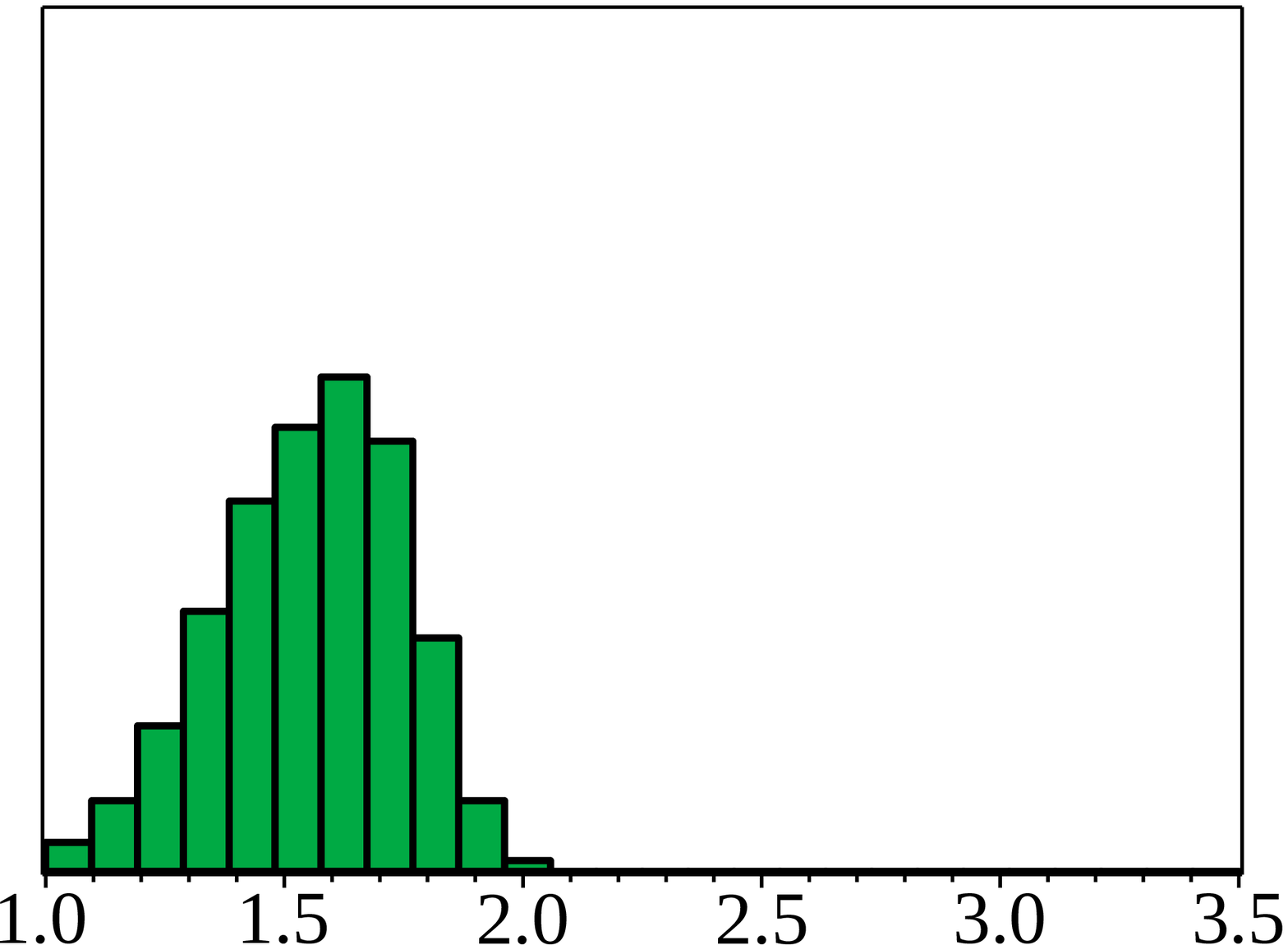}
        \caption{$l_\mathrm{c}=0.25$.}
    \end{subfigure}\\
    \caption{The frequency histograms of velocity variations due to impact of the correlation length $l_\mathrm{c}$ for $\delta_\mathrm{d}=\pm 0.03 k_0$.}
    \label{fig:hyst_lc}
\end{figure}

The typical water-polymer distributions obtained for the permeability maps in Figure\,\ref{fig:perm_lc} as input parameters for numerical simulations are presented in Figure\,\ref{fig:fingers_lc}. The velocity of the fastest finger tip was determined by linear regression of the $x_\mathrm{f}(t)$ curves on the time interval 0.025-0.55\,PV, assuming a constant growth rate of viscous fingers and no influence of numerical diffusion, as in\,\cite{Linear,koval1963}. In order to obtain the distribution density of the fingers velocity, the resulting statistics were subjected to a frequency analysis in the intervals from 1 to 3.5\,PV$^{-1}$ with a step of 0.4\,PV$^{-1}$, followed by normalisation by the number of trials (see, Fig.\,\ref{fig:hyst_lc}). After a direct comparison of the distribution density curves for fine-, medium-, and coarse-grained formation structures, it can be concluded that an increase in coarseness first leads to a slight expansion of the distribution shape and a shift of the maximum velocity to the large velocity. A further increase in the correlation length results in a reversal of the behaviour.

Box diagrams are useful for visualizing the discovered statistical distributions. This type of chart shows in a simple way shows the median, lower, and upper quartiles (50$^\mathrm{th}$, 25$^\mathrm{th}$, and 75$^\mathrm{th}$ percentiles, respectively), as well as the minimum and maximum values of the sample statistics. The distances between different parts of the graph allow determining the degree of scatter (dispersion) and asymmetry of the data. Using a fine-grained formation structure as an example ($l_\mathrm{c}=5\cdot 10^{-3}$), the third quartile or the 75$^\mathrm{th}$ percentile of the growth rate of viscous fingers is 1.73\,PV$^{-1}$. This means that 75\% of the fingers grow at a rate less than or equal to 1.73\,PV$^{-1}$.

\begin{figure}[t]
    \centering
     \includegraphics[width=0.55\textwidth]{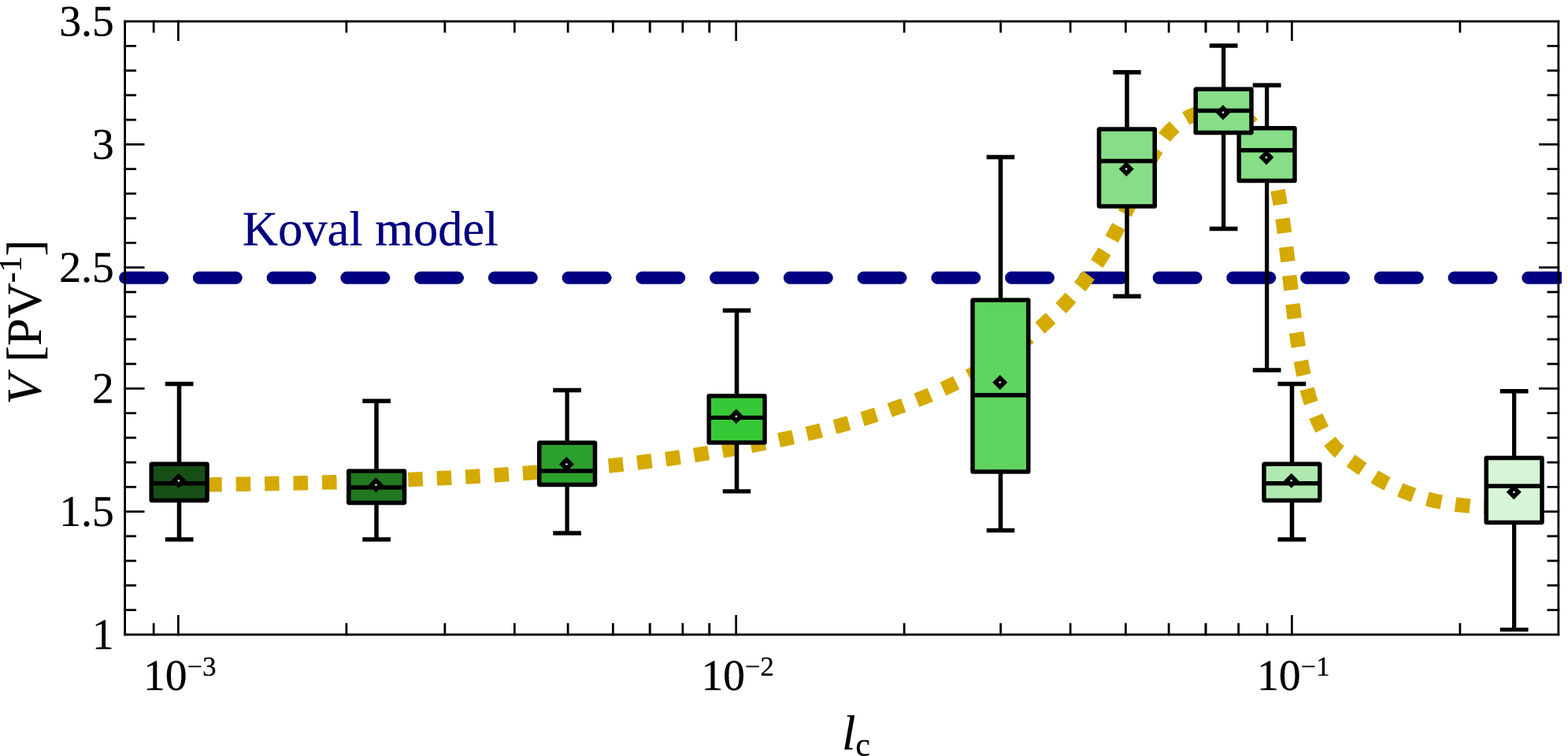}
        \caption{Box diagrams illustrating the statistical distribution of viscous finger velocity for a range of reservoir permeability correlation lengths for $\delta_\mathrm{d}=\pm 0.03 k_0$. The dotted line corresponds to the moving average of the mean data sets. The Koval model estimate of $V=2.46$\,PV$^{-1}$ is depicted for reference (dashed line).}
\label{fig:lc_stats}  
\end{figure}

\begin{figure}[t!]
    \centering
    \begin{subfigure}{0.3\textwidth}
        \includegraphics[height=0.15\textheight]{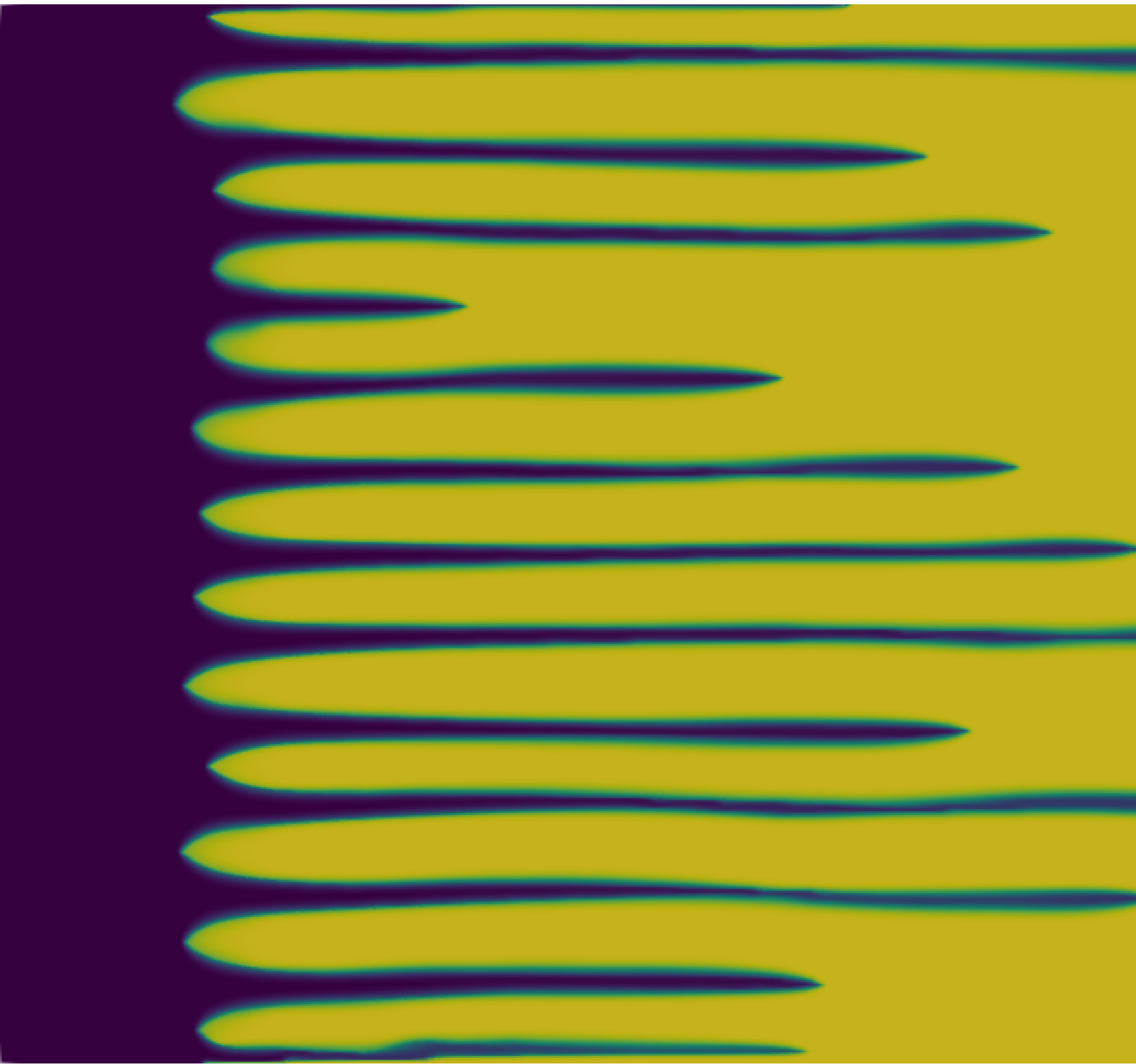}
        \caption{}
    \end{subfigure}
    \begin{subfigure}{0.3\textwidth}
        \includegraphics[height=0.15\textheight]{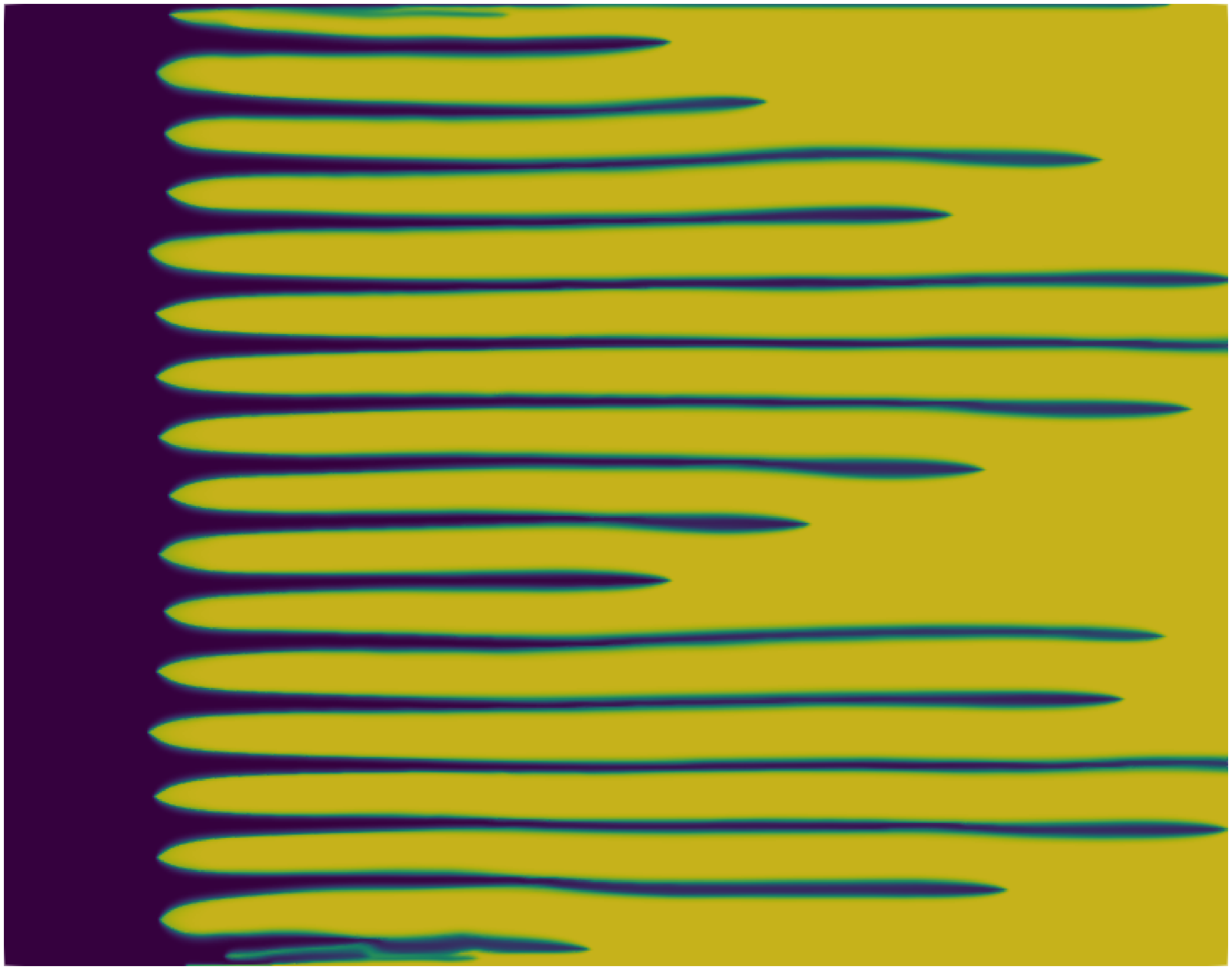}
        \caption{}
    \end{subfigure}
    \begin{subfigure}{0.3\textwidth}
        \includegraphics[height=0.15\textheight]{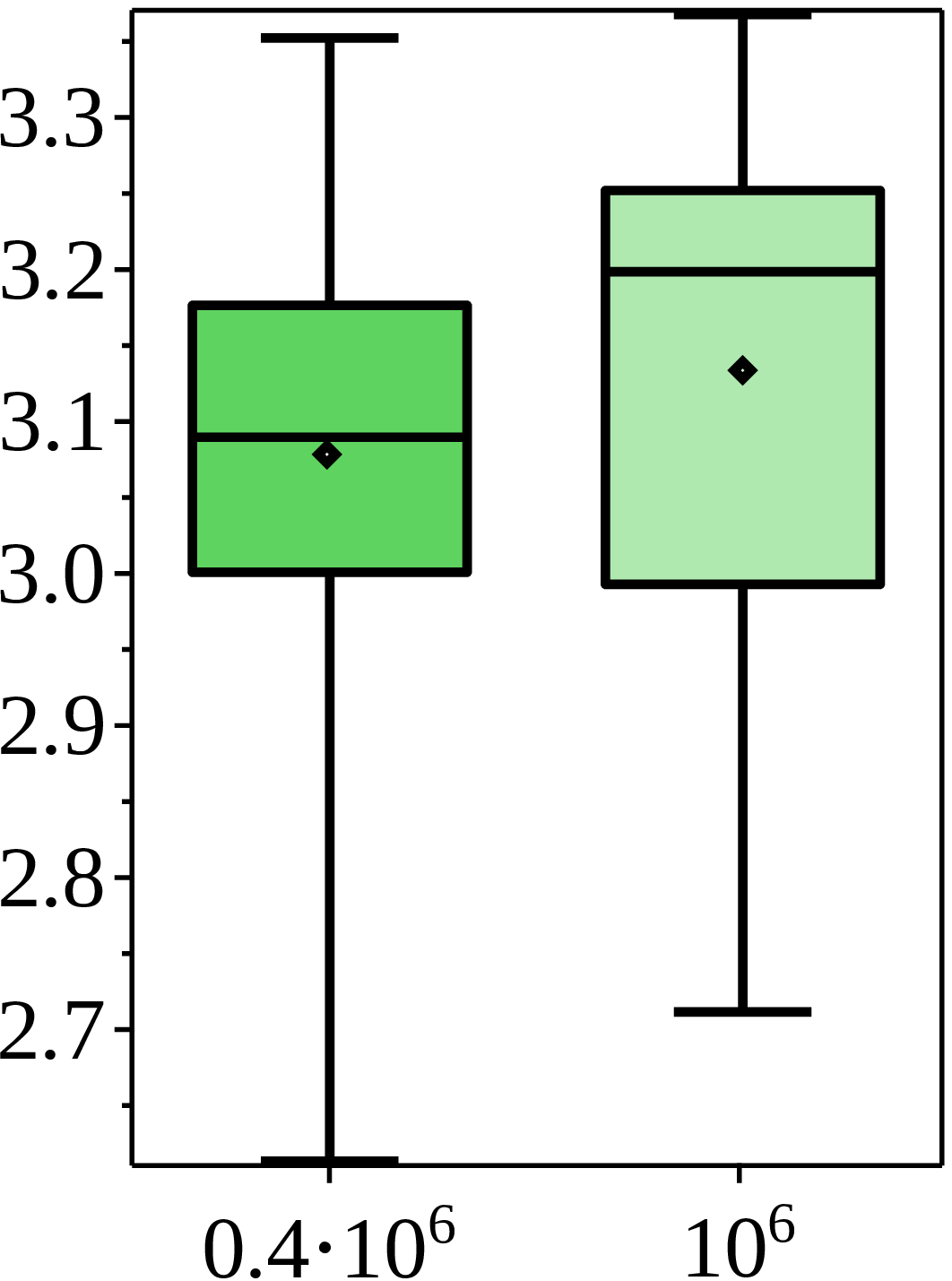}
        \caption{}
    \end{subfigure}\\
        \begin{subfigure}{0.3\textwidth}
        \includegraphics[height=0.15\textheight]{Figures/fingers_100_010.eps}
        \caption{}
    \end{subfigure}
    \begin{subfigure}{0.3\textwidth}
        \includegraphics[height=0.15\textheight]{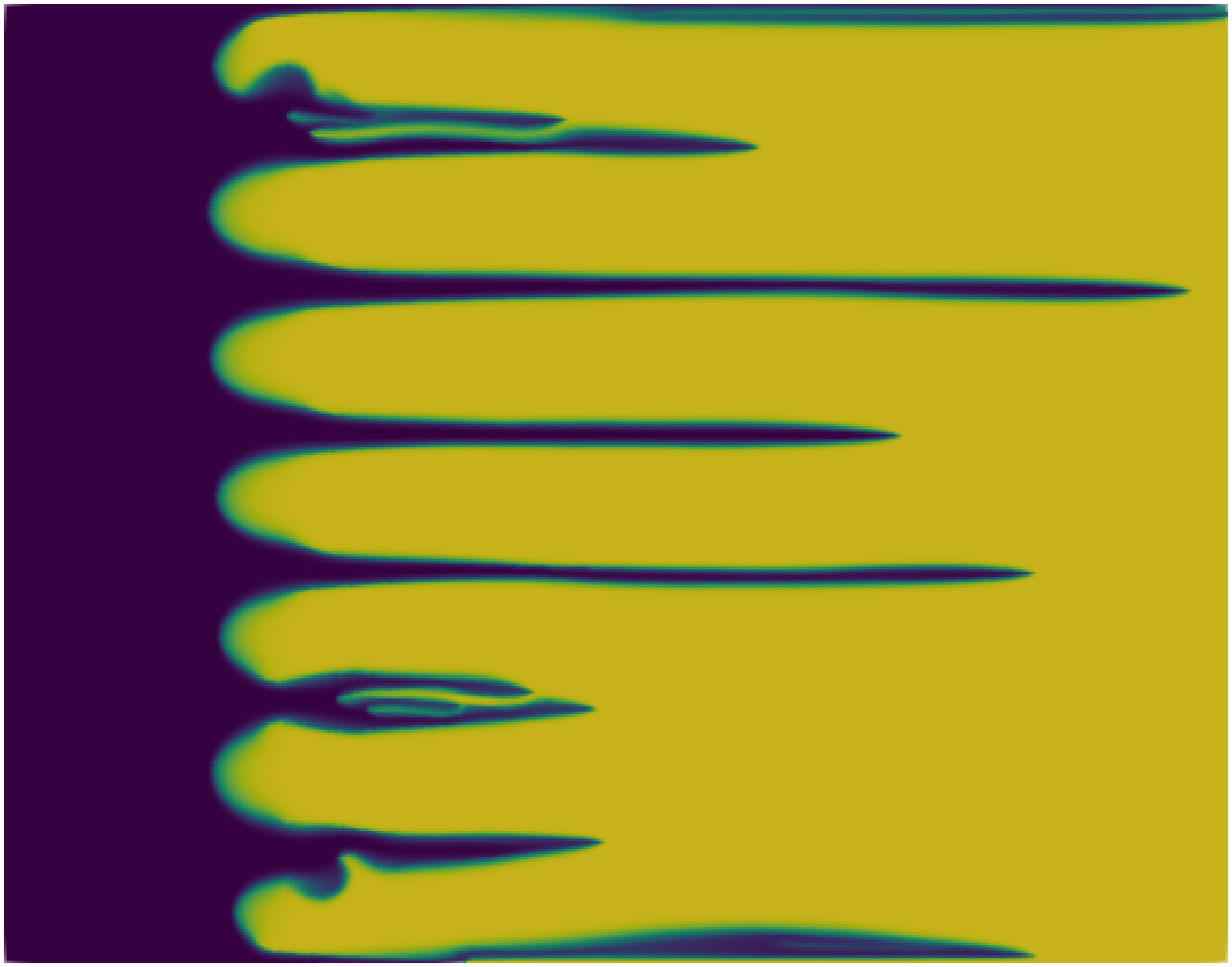}
        \caption{}
    \end{subfigure}
    \begin{subfigure}{0.3\textwidth}
        \includegraphics[height=0.15\textheight]{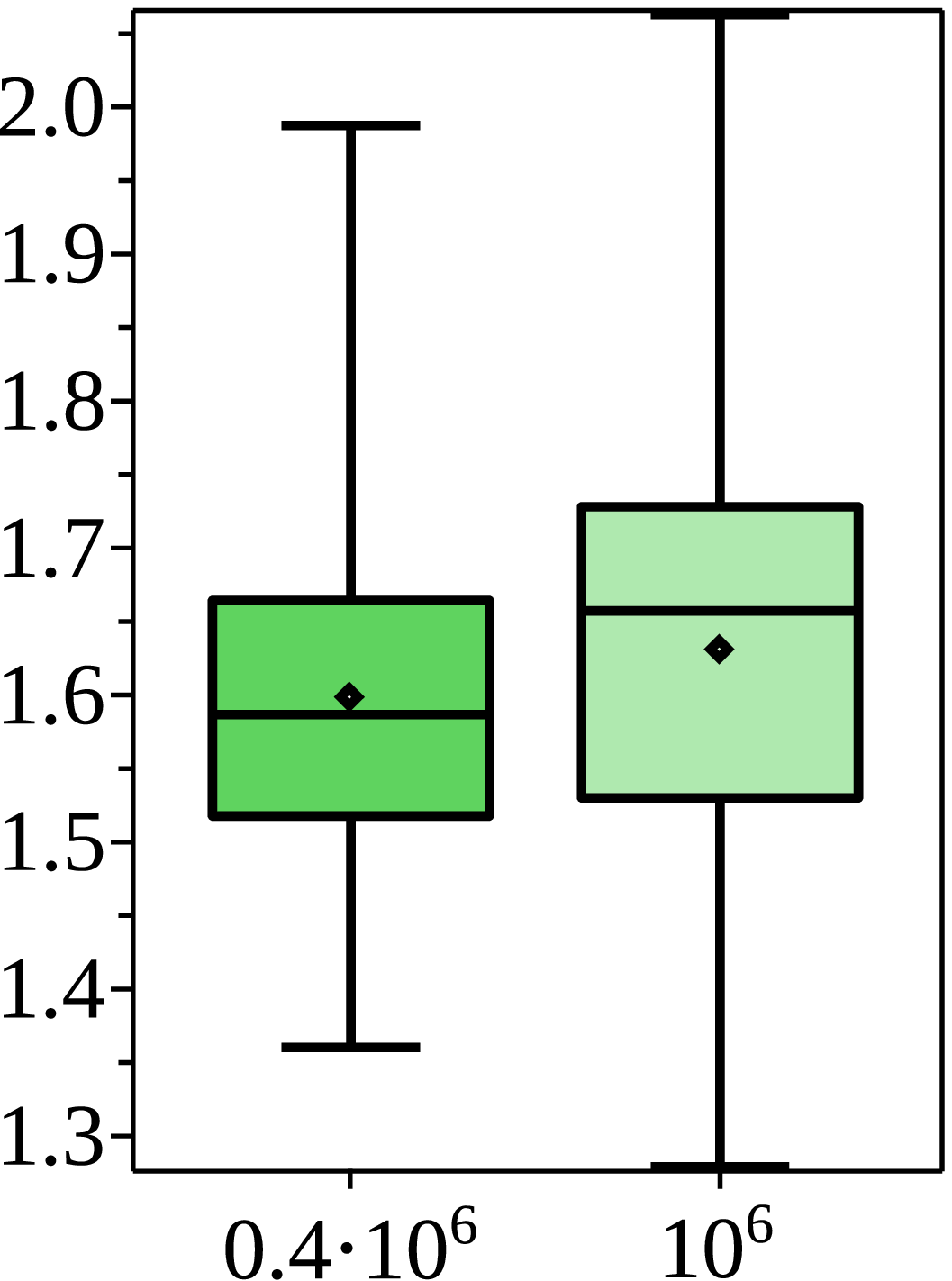}
        \caption{}
    \end{subfigure}
    \caption{The polymer profiles computed on the grid with $0.4\cdot10^6$ elements (a) and with $10^6$ elements (b) for $l_\mathrm{c}=7.5\cdot10^{-2}$. (c) A comparison of the viscous fingers statistics for the different computational grids and reservoir permeability with a correlation length of $l_\mathrm{c}=7.5\cdot10^{-2}$ and a dispersion of $\delta_\mathrm{d}=\pm 0.1k_0$. The sub-figures (d),(e),(f) are identical to the sub-figures (a),(b),(c) with the exception of $l_\mathrm{c}=1\cdot 10^{-1}$.
    }
    \label{fig:verify}
\end{figure}

For the verification of the obtained statistical distributions, we repeated a series of simulations on a computational grid with smaller elements. In this case, the total number of elements was about $10^6$ with a grid cell size of 3.6\,cm. Figure\,\ref{fig:verify} compares polymer profiles and box diagrams for the fingers velocities at a correlation length of $l_\mathrm{c}=7.5\cdot10^{-2}$ and $l_\mathrm{c}=1\cdot10^{-1}$.

An increase in the number of grid elements does not significantly affect the results and leads to a slight thinning of the width of the viscous fingers, inducing a decrease in the statistical spread of their velocities. Such a behavior is explained by a decrease in numerical diffusion and lower blurring of the front with a larger number of grid elements.

\section{Influence of the dispersion of reservoir permeability}
\label{sec:ds}

\begin{figure}[t]
    \centering
     \includegraphics[width=0.49\textwidth]{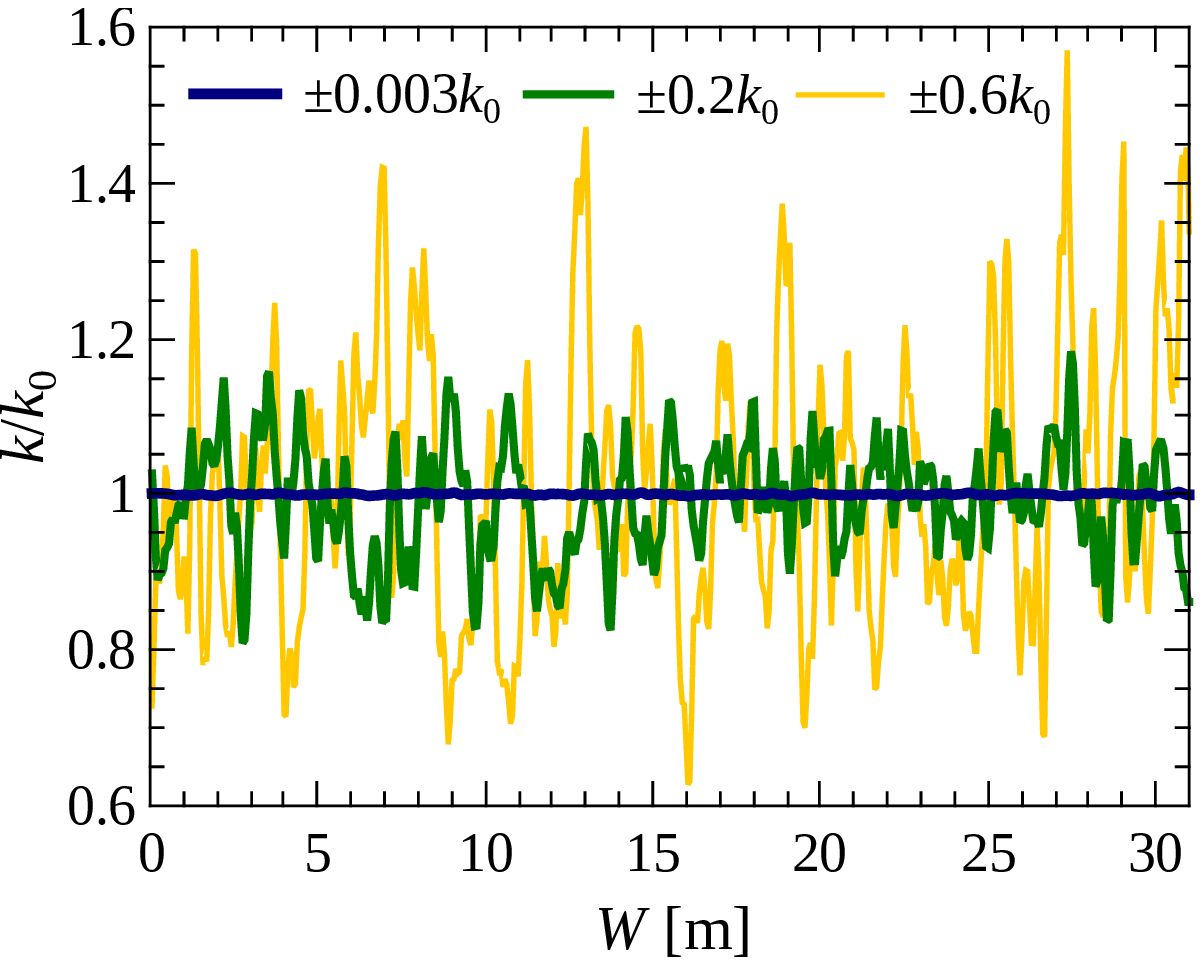}
        \caption{The coordinate profiles of the reservoir permeability for different dispersion characteristics and $l_\mathrm{c}=5\cdot10^{-3}$.}
\label{fig:perm_ld}  
\end{figure}

We applied similar analysis to reveal the effect of the standard deviation of the mean value of reservoir permeability on the growth of viscous fingers through the polymer bank. The medium-grained formation structure with $l_\mathrm{c}=5\cdot10^{-3}$ was chosen for these studies. The coordinate profiles of reservoir permeability for dispersion values $\delta_\mathrm{d}=\pm0.003k_0$, $\pm0.2k_0$, and $\pm0.6k_0$ are shown in Fig.\,\ref{fig:perm_ld}.

\begin{figure}[t!]
    \centering
    \begin{subfigure}{0.32\textwidth}
        \includegraphics[height=0.17\textheight]{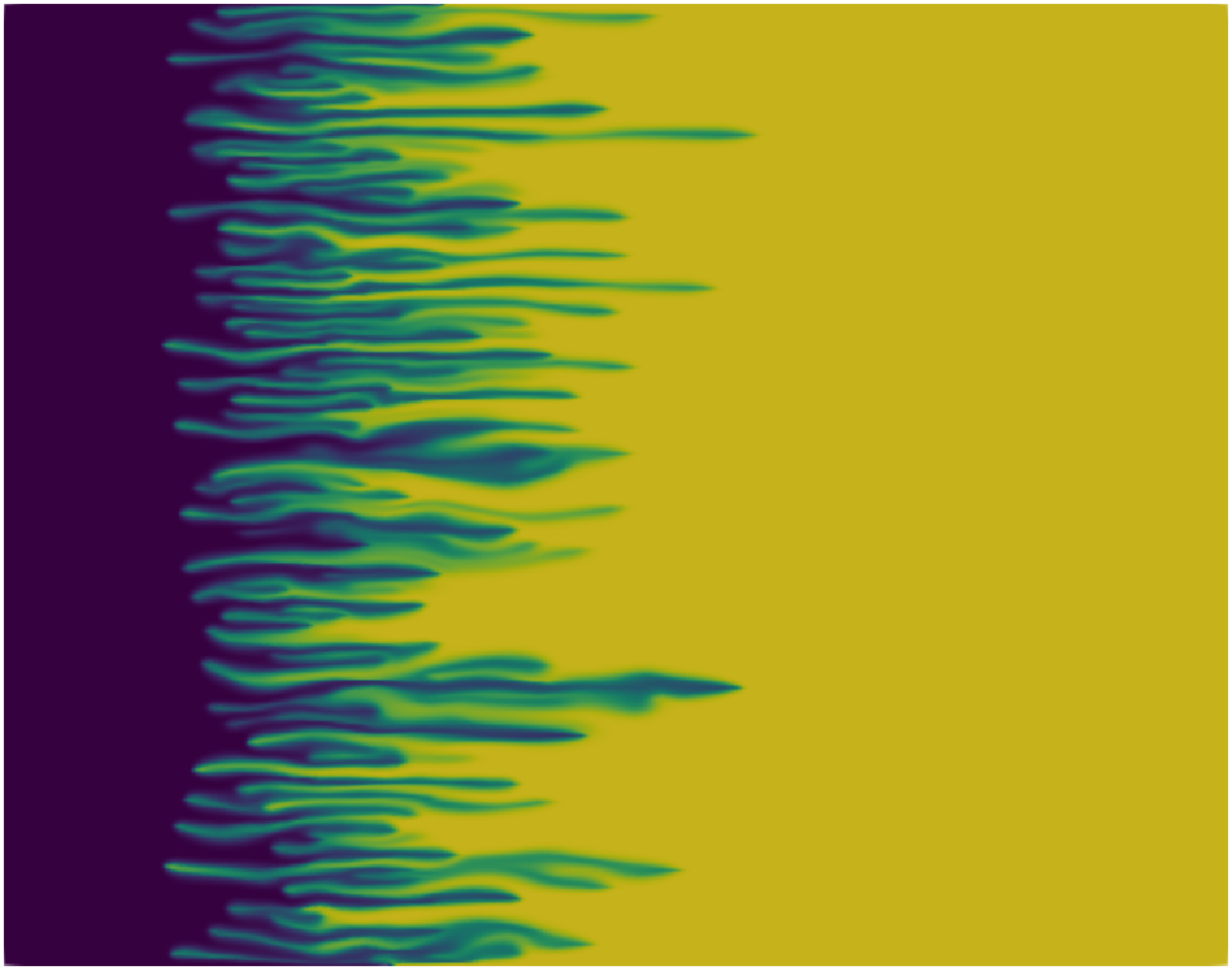}
        \caption{$\delta_\mathrm{d}=\pm10^{-3}$.}
    \end{subfigure}
    \begin{subfigure}{0.32\textwidth}
        \includegraphics[height=0.17\textheight]{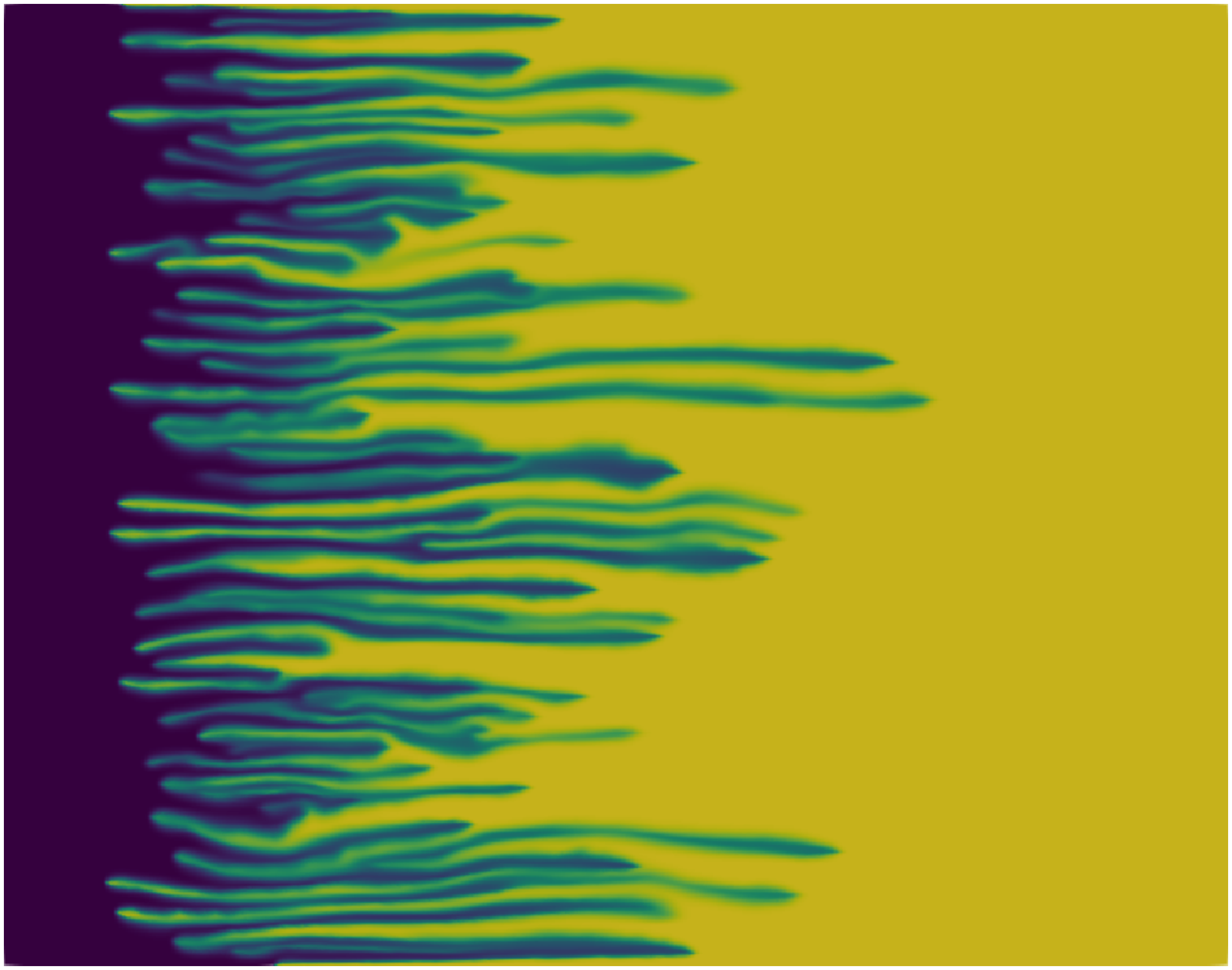}
        \caption{$\delta_\mathrm{d}=\pm7\cdot10^{-2}$.}
    \end{subfigure}
    \begin{subfigure}{0.32\textwidth}
        \includegraphics[height=0.17\textheight]{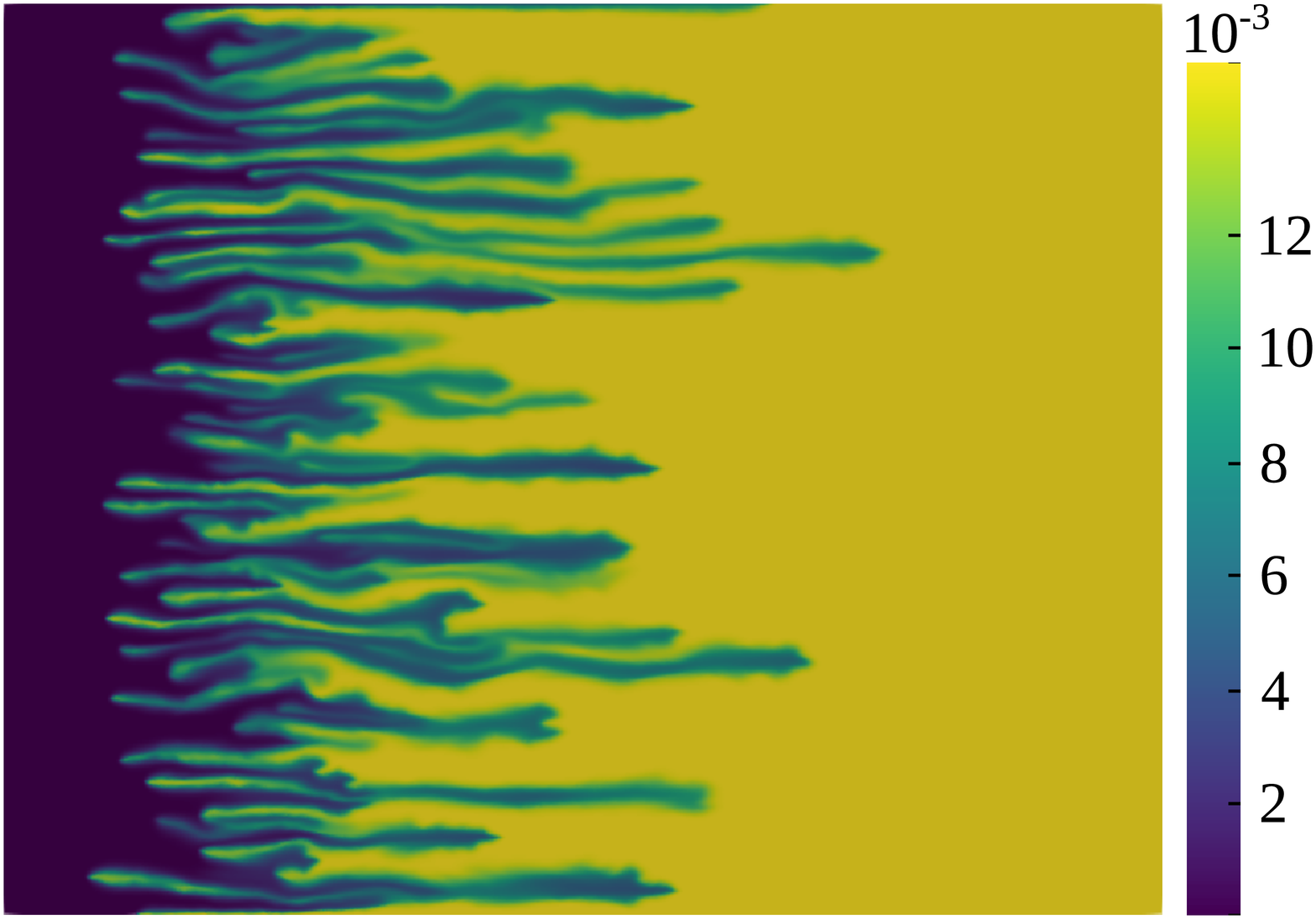}
        \caption{$\delta_\mathrm{d}=\pm0.15$.}
    \end{subfigure}
    \caption{The viscous fingers in the polymer flood formed after 0.4\,PV water injection for a set of standard deviations of the mean value of reservoir permeability, as shown in Figure\,\ref{fig:perm_ld}.}
    \label{fig:fingers_ld}
\end{figure}

\begin{figure}[t!]
    \centering
    \begin{subfigure}{0.3\textwidth}
        \includegraphics[height=0.15\textheight]{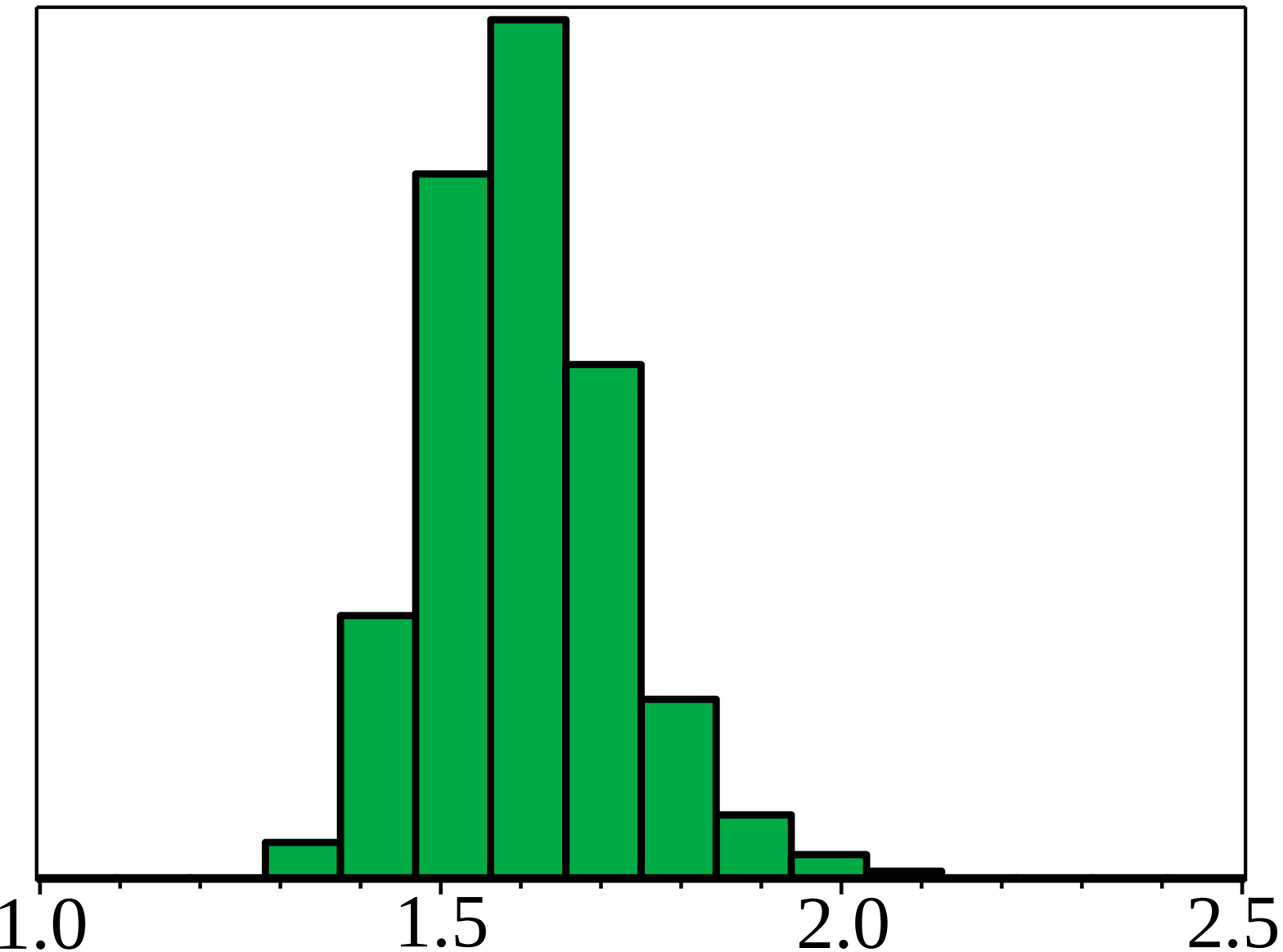}
        \caption{$\delta_\mathrm{d}=\pm10^{-3}$.}
    \end{subfigure}
    \begin{subfigure}{0.3\textwidth}
        \includegraphics[height=0.15\textheight]{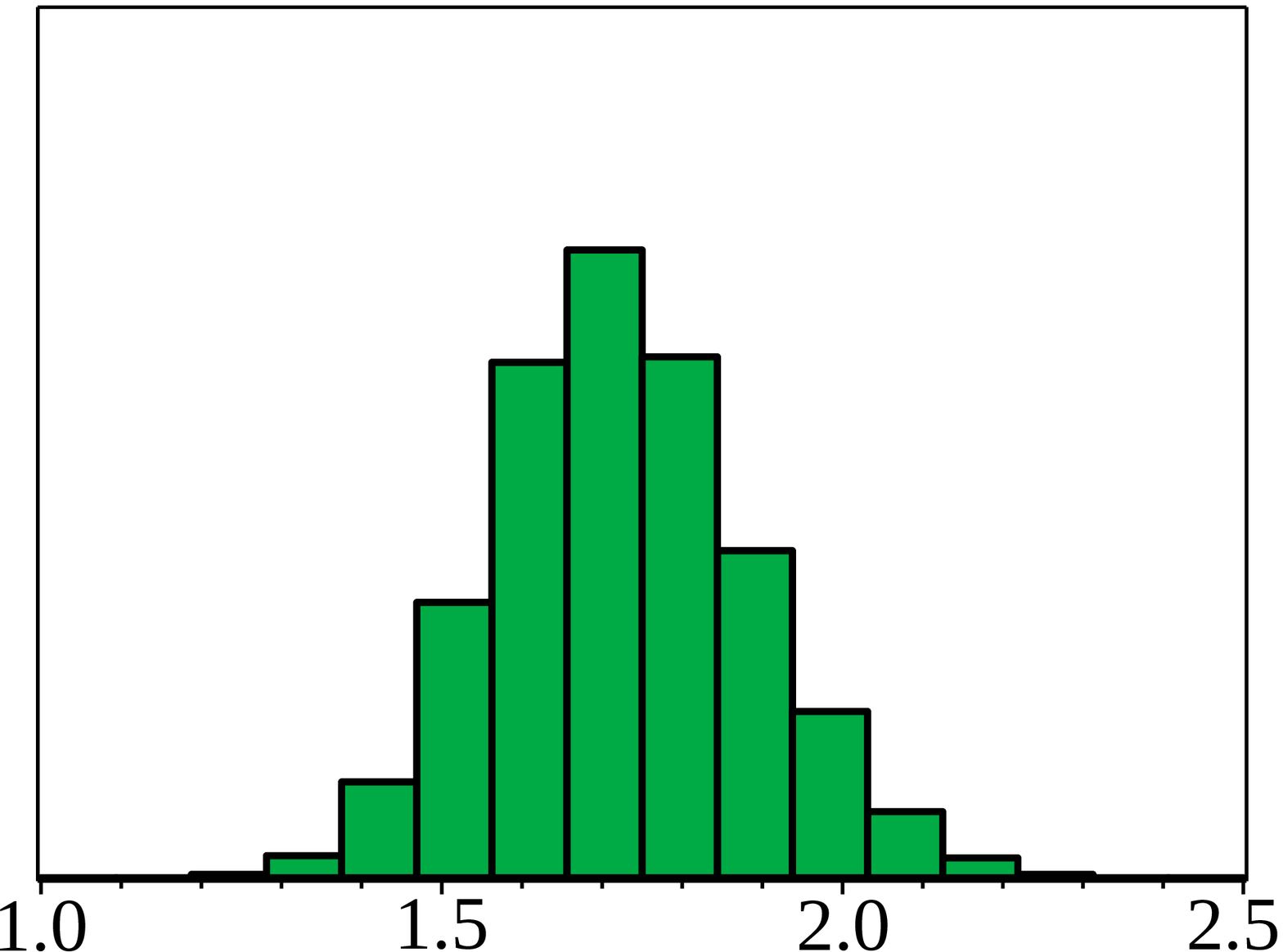}
        \caption{$\delta_\mathrm{d}=\pm7\cdot10^{-2}$.}
    \end{subfigure}
    \begin{subfigure}{0.3\textwidth}
        \includegraphics[height=0.15\textheight]{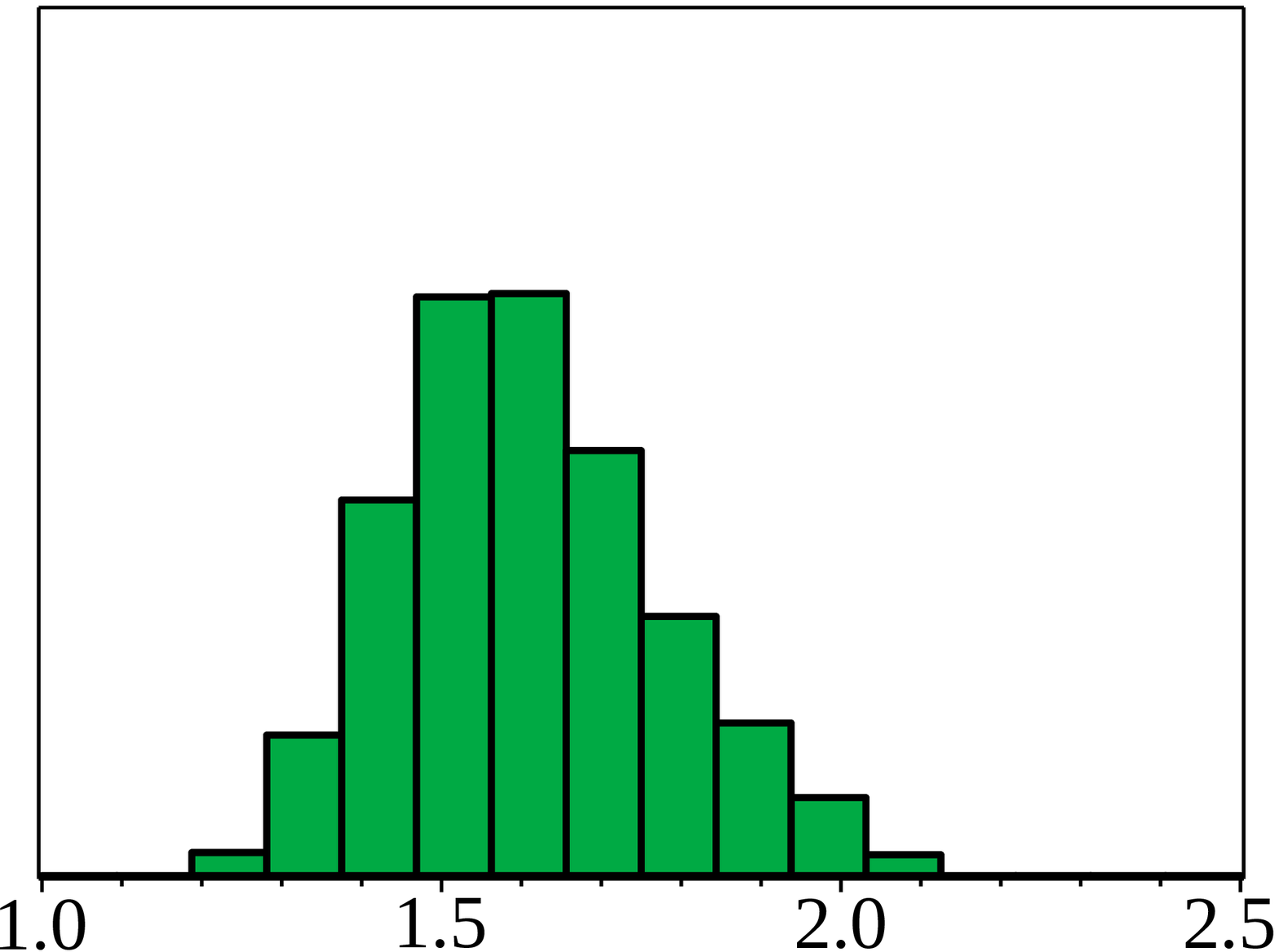}
        \caption{$\delta_\mathrm{d}=\pm0.15$.}
    \end{subfigure}
    \caption{The frequency histograms of the velocity fluctuations due to the impact of the standard deviations of the mean value of reservoir permeability for $l_\mathrm{c}=5\cdot10^{-3}$.}
    \label{fig:hyst_ld}
\end{figure}

\begin{figure}[t]
    \centering
     \includegraphics[width=0.49\textwidth]{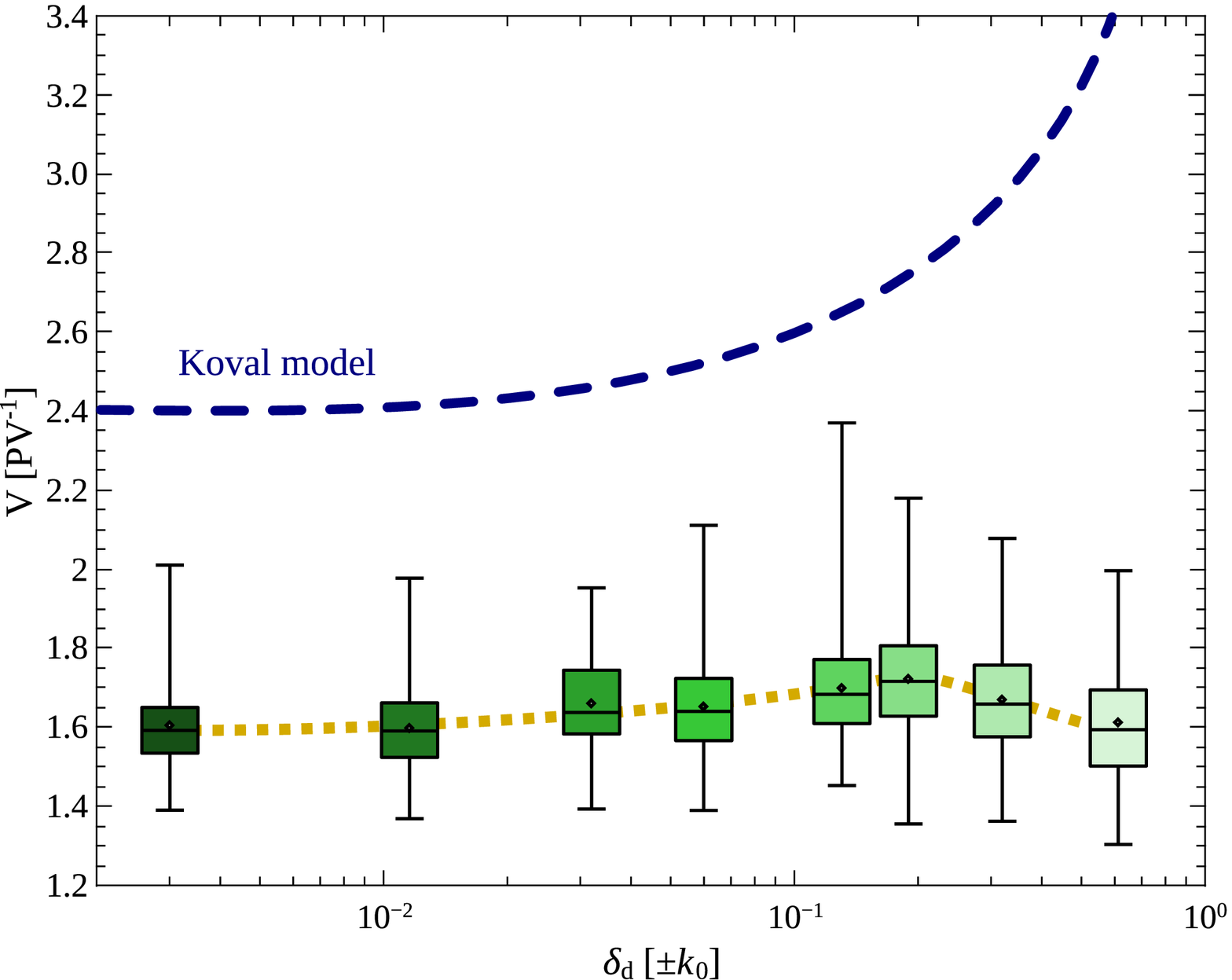}
        \caption{Box diagrams displaying the statistical distribution of viscous fingers for a given set of standard deviations from the mean value of reservoir permeability for $l_\mathrm{c}=5\cdot10^{-3}$. The dotted line corresponds to the moving average of the mean data-sets and the dashed line to the Koval estimates.}
\label{fig:ld_stats}  
\end{figure}

Similarly, statistics were obtained on the growth of viscous fingers in the water-polymer system at a range of deviations from the mean value of $k_0$ using a sample of 200 different permeability profiles at $l_\mathrm{c}=5\cdot10^{-3}$. Figures\,\ref{fig:hyst_ld} and \ref{fig:ld_stats} demonstrate that the shape and properties of the distribution density change slightly when the standard deviation of the permeability does not exceed a few percent of $k_0$. Such an insignificant change on the order of 2\% may be a statistical error due to the peculiarities of the modeling process and processing of the results. Nonetheless, similar to the dependence on the correlation length, a clear maximum of the velocity dependence on the magnitude of the dispersion can be distinguished at $l_\mathrm{d}=\pm 0.19 k_0$.

\section{Results and discussion}
From the observed statistics, we may conclude that the behavior of viscous fingers is rather chaotic at low values of graininess. Above a certain value of the correlation length, conductive channels of various widths begin to form in the reservoir. This contributes to an increase in the growth rate of the viscous fingers. Similar behaviour was observed within a graph-based model of immiscible displacement\,\cite{master2003}. As the correlation length increases further, the number of conductive channels in the reservoir begins to decrease. As a result, the formation and growth of fingers slow down dramatically.

It is interesting to compare the numerical results to analytical estimates by the transverse flow equilibrium (TFE) approximation (Otto-Menon-Yortsos-Salin model\,\cite{otto2006,yortsos2006}) and the Koval empirical approach\,\cite{koval1963}. Following these two models, we can calculate the velocity of front end of the mixing zone $v$ as

\begin{equation}
\begin{aligned}
v^\mathrm{TFE}=\mu(c_\mathrm{max})H^\mathrm{TFE}\int\limits_0^{c_\mathrm{max}}\cfrac{\mathrm{d} c}{\mu(c)},\\
v^\mathrm{Koval}=H^\mathrm{Koval}\left[a M^{1/4}+(1-a)\right]^4,
\label{eq:estimates}
\end{aligned}
\end{equation}
where $M=\mu(c_\mathrm{max})/\mu(0)=20$ defines the ratio of the maximum and minimum values of water viscosity. In the Koval model, the parameter $a$ is typically set to 0.22\,\cite{Linear,koval1963}. The factors $H^\mathrm{TFE,Koval}$ characterize the permeability heterogeneity of the reservoir. In the TFE approximation the ratio $\max(k(x,y))/k_0$ defines $H_\mathrm{TFE}$ for the front velocity of the mixing zone\,\cite{yortsos2006}. The Koval heterogeneity factor is empirically related to the Dykstra-Parsons coefficient $V_\mathrm{DP}$ as\,\cite{brantson2018,arya1986} $\log_{10} \left(H^\mathrm{Koval}\right) = V_\mathrm{DP} \left(1-V_\mathrm{DP}\right)^{-0.2}$. The $V_\mathrm{DP}$ coefficient is the most commonly used static measure of permeability variation and is given by $V_\mathrm{DP}=\left(\hat k-k_\sigma\right)/\hat k$, where $\hat k$ is the median permeability and $k_\sigma$ is the one standard deviation of the cumulative sample\,\cite{koval1963,brantson2018}.

Thus, the velocity of the front boundary of the mixing zone is estimated as $v^\mathrm{TFE} =7.03$\,PV$^{-1}$ by TFE and $v^\mathrm{Koval} =2.46$\,PV$^{-1}$ by the Koval model for dispersion $l_\mathrm{d}=\pm 0.03k_0$. This is consistent with the findings of\,\cite{Linear} for a homogeneous reservoir, which show that the Koval approach is more accurate for the quadratic viscosity and the velocity of viscous fingers is significantly more pessimistic under the transverse flow equilibrium approximation.

Although the influence of correlation length is not considered in the aforementioned estimates, the calculated value of finger growth rate for fine-grained reservoir permeability proved to be lower than the mean values for medium- and coarse-grained structures in Fig.\,\ref{fig:lc_stats}. In turn, the Koval estimates of growth rate nearly agree with the mean value of the velocity range as a function of reservoir structure heterogeneity $l_\mathrm{c}$ (2.46\,PV$^{-1}$ versus the interval from about 1.6\,PV$^{-1}$ to 3.4\,PV$^{-1}$). Meanwhile, the analytical curve in Fig.\,\ref{fig:fingers_ld} appears somewhat pessimistic even according to the Koval model. This may confirm the view that the Koval estimates for heterogeneous media are valid for large correlation lengths~\cite{arya1986}. From observation of the constructed graphs, it can be concluded that both approaches \eqref{eq:estimates} do not reproduce the non-monotonic nature of the growth rate of the fingers.

\section{Conclusions}
\label{sec:summary}
On the basis of the implemented numerical scheme and performed experiments, the distribution functions of the velocities of viscous fingers have been obtained for the fine-, medium-, and coarse-grained structures of the reservoir permeability. It is shown that an increase in heterogeneity (graininess) leads to a non-monotonic dependence of the growth rate of the fingers. There exists a maximum growth rate of fingers at certain parameters of the medium granularity. Depending on the nature of permeability heterogeneity, the difference in velocities of viscous fingers can be as much as a factor of two. The theoretical estimates based on the Koval approach for the velocity of the tip of the leading finger fall within the correct range of the growth rates and are not pessimistic for the quadratic viscosity function. This implies that the velocity of the front part of the water-polymer mixing zone at a small coarseness turns out to be underestimated compared to the mean values for the medium- and coarse-grained reservoir structures. These reported statistics can be used to create a probabilistic model for the breakthrough of a slug of fluids of different viscosities when using chemical methods to enhance oil production.

\section*{Acknowledgements}
\noindent
The work of F.L.~Bakharev, D.A.~Pavlov, and I.A.~Starkov was performed at the Saint-Petersburg Leonhard Euler International Mathematical Institute and supported by the Ministry of Science and Higher Education of the Russian Federation (agreement no.~075-15-2022-287).

\end{document}